\documentclass[conference]{IEEEtran}
\IEEEoverridecommandlockouts
\usepackage{amsmath,amssymb}
\usepackage{pgfplotstable}
\usepgfplotslibrary{colormaps}
\usepackage{subcaption}
\usepackage{siunitx}
\usepackage{url}
\definecolor{C0}{RGB}{031, 119, 180} 
\definecolor{C1}{RGB}{255, 127, 014} 
\definecolor{C2}{RGB}{044, 160, 044} 
\definecolor{C3}{RGB}{215, 039, 040} 
\definecolor{C4}{RGB}{148, 103, 189} 
\definecolor{C5}{RGB}{140, 086, 075} 
\definecolor{C6}{RGB}{227, 119, 194} 
\definecolor{C7}{RGB}{127, 127, 127} 
\definecolor{C8}{RGB}{188, 189, 034} 
\definecolor{C9}{RGB}{023, 190, 207} 

\definecolor{C10}{RGB}{174, 199, 232} 
\definecolor{C11}{RGB}{255, 187, 120} 
\definecolor{C12}{RGB}{152, 223, 138} 
\definecolor{C13}{RGB}{255, 152, 150} 
\definecolor{C14}{RGB}{197, 176, 213} 
\definecolor{C15}{RGB}{196, 156, 148} 
\definecolor{C16}{RGB}{247, 182, 210} 
\definecolor{C17}{RGB}{199, 199, 199} 
\definecolor{C18}{RGB}{219, 219, 141} 
\definecolor{C19}{RGB}{158, 218, 229} 

\pgfplotsset{
  colormap={cmaporange}{
    rgb255=(255,255,255)
    rgb255=(255, 127, 014)
  }
}

\pgfplotsset{
  colormap={cmapblue}{
    rgb255=(255,255,255)
    rgb255=(031, 119, 180)
  }
}

\pgfplotsset{
  colormap={cmapgreen}{
    rgb255=(255,255,255)
    rgb255=(044, 160, 044)
  }
}
\usepackage[font=small]{caption}
 \usetikzlibrary{backgrounds}
\usepackage{import}
\usepackage{titlesec}
\usepackage{tikz}
\usepackage{import}

\usetikzlibrary{positioning}
\usetikzlibrary{fit}
\usetikzlibrary{calc}
\usetikzlibrary{shapes}
\usetikzlibrary{math}
\usetikzlibrary{fpu}
\usetikzlibrary{decorations.markings}
\usetikzlibrary{decorations.pathreplacing}
\usetikzlibrary{arrows.meta}
\usetikzlibrary{intersections}

\makeatletter

\tikzset{outer sep=0}
\tikzset{inner sep=0}

\def\pgfaddtoshape#1#2{
	\begingroup
	\def\pgf@sm@shape@name{#1}%
	\let\anchor\pgf@sh@anchor
	#2%
	\endgroup
}

\newcommand{\anchorlet}[2]{
	\global\expandafter
	\let\csname pgf@anchor@\pgf@sm@shape@name @#1\expandafter\endcsname
	\csname pgf@anchor@\pgf@sm@shape@name @#2\endcsname
}

\pgfmathsetmacro{\NODESIZE}{42}
\pgfmathsetmacro{\NODETHICKNESS}{1.0}
\pgfmathsetmacro{\ROUNDEDCORNERS}{0.5mm}
\tikzset{node distance=0.5*\NODESIZE pt}

\def\FNODESIZE{\NODESIZE pt}

\tikzstyle{textstyle} = [text height=1.5ex, text depth=.5ex]
\tikzset{every label/.style=textstyle}

\tikzstyle{linestyle} = [line width = \NODETHICKNESS, rounded corners = \ROUNDEDCORNERS]
\tikzstyle{arrowstyle} = [>=stealth, linestyle]
\tikzstyle{<--} = [<-, arrowstyle]
\tikzstyle{-->} = [->, arrowstyle]
\tikzstyle{->-} = [linestyle, decoration={markings,	mark=at position 0.5 with {\arrow[arrowstyle]{>}}}, postaction={decorate}] 
\tikzstyle{-<-} = [linestyle, decoration={markings,	mark=at position 0.5 with {\arrow[arrowstyle]{<}}}, postaction={decorate}] 

\tikzset{   
    -A-/.style args={#1}{%
        linestyle, 
        decoration={markings, mark=at position #1 with {\arrow[>=Triangle, scale=.025*\NODESIZE]{>}}},
        postaction={decorate}
    },
    -A-/.default = {0.75}
}

\tikzset{   
    -AA-/.style args={#1, #2, #3}{%
        linestyle,
        decoration={markings, mark=at position #1 with {
            \node[amp={color=#2, width=#3, height=#3}, rotate=\pgfdecoratedangle] at (0,0) (inline_amp) {};
        }},
        postaction={decorate}
    },
    -AA-/.default = {0.5, E, 0.15}
}

\tikzstyle{<-->} = [<->, arrowstyle]
\tikzstyle{---} = [arrowstyle]

\tikzset{
    -PC-/.style args={#1}{
        linestyle, 
        decoration={markings, mark=at position #1 with {
            \draw[---,FO]
                (.05*\NODESIZE*\pgflinewidth,.05*\NODESIZE*\pgflinewidth) circle[radius=.05*\NODESIZE*\pgflinewidth];
            \draw[---,FO]
                (-.05*\NODESIZE*\pgflinewidth,.05*\NODESIZE*\pgflinewidth) circle[radius=.05*\NODESIZE*\pgflinewidth];
            \draw[---,FO]
                (0,-.05*\NODESIZE*\pgflinewidth) circle[radius=.05*\NODESIZE*\pgflinewidth];}},
        postaction={decorate}
        },
        -PC-/.default = {0.5}
}

\definecolor{C0}{RGB}{031, 119, 180} 
\definecolor{C1}{RGB}{031, 119, 180} 
\definecolor{C2}{RGB}{255, 127, 014} 
\definecolor{C3}{RGB}{044, 160, 044} 
\definecolor{C4}{RGB}{215, 039, 040} 
\definecolor{C6}{RGB}{148, 103, 189} 
\definecolor{C100}{RGB}{140, 086, 075} 
\definecolor{C7}{RGB}{227, 119, 194} 
\definecolor{C8}{RGB}{127, 127, 127} 
\definecolor{C9}{RGB}{188, 189, 034} 
\definecolor{C5}{RGB}{023, 190, 207} 

\definecolor{C0l}{RGB}{174, 199, 232} 
\definecolor{C11}{RGB}{174, 199, 232} 
\definecolor{C12}{RGB}{255, 187, 120} 
\definecolor{C13}{RGB}{152, 223, 138} 
\definecolor{C14}{RGB}{255, 152, 150} 
\definecolor{C15}{RGB}{197, 176, 213} 
\definecolor{C16}{RGB}{196, 156, 148} 
\definecolor{C17}{RGB}{247, 182, 210} 
\definecolor{C18}{RGB}{199, 199, 199} 
\definecolor{C19}{RGB}{219, 219, 141} 
\definecolor{C20}{RGB}{158, 218, 229} 

\definecolor{Snow}{HTML}{FBFBFB} 			
\definecolor{TUeRed}{RGB}{200, 25, 25}		
\definecolor{TUeGreen}{RGB}{25, 200, 113}	
\definecolor{TUeBlue}{RGB}{25, 113, 200}	

\definecolor{O}{RGB}{031, 119, 180} 	
\definecolor{Ol}{RGB}{174, 199, 232} 	
\definecolor{E}{RGB}{255, 127, 014} 	
\definecolor{El}{RGB}{255, 187, 120} 	
\definecolor{D}{RGB}{148, 103, 189}     
\definecolor{Dl}{RGB}{197, 176, 213} 	
\definecolor{EO}{RGB}{215, 039, 040} 	
\definecolor{EOl}{RGB}{255, 152, 150}   

\tikzstyle{FW} = [fill=white]			
\tikzstyle{FB} = [fill=white]			
\tikzstyle{FO} = [fill=C0l, draw=C0]	
\tikzstyle{FE} = [fill=C1l, draw=C1]	
\tikzstyle{FD} = [fill=C2l, draw=C2]	

\def\direce{e}
\def\direcw{w}
\def\direcn{n}
\def\direcs{s}

\def\flipfalse{0}
\newcount\portcount
\pgfdeclareshape{basic}{
	\inheritsavedanchors[from=rectangle] 
	
	\inheritanchor[from=rectangle]{center}
	\inheritanchor[from=rectangle]{mid}		
	\inheritanchor[from=rectangle]{base}	
	\inheritanchor[from=rectangle]{north}
	\inheritanchor[from=rectangle]{south}		
	\inheritanchor[from=rectangle]{west}		
	\inheritanchor[from=rectangle]{mid west}				
	\inheritanchor[from=rectangle]{base west}		
	\inheritanchor[from=rectangle]{north west}		
	\inheritanchor[from=rectangle]{south west}		
	\inheritanchor[from=rectangle]{east}
	\inheritanchor[from=rectangle]{mid east}
	\inheritanchor[from=rectangle]{base east}	
	\inheritanchor[from=rectangle]{north east}
	\inheritanchor[from=rectangle]{south east}	
	
	\inheritanchorborder[from=rectangle]	

	\inheritbackgroundpath[from=rectangle]	
}

\pgfaddtoshape{basic}{
	\anchor{west south west}{
		\pgf@process{\northeast}
		\pgf@ya=.5\pgf@y
		\pgf@process{\southwest}
		\pgf@y=1.5\pgf@y
		\advance\pgf@y by \pgf@ya
		\pgf@y=.5\pgf@y
	}
	\anchor{west north west}{
		\pgf@process{\northeast}
		\pgf@ya=1.5\pgf@y
		\pgf@process{\southwest}
		\pgf@y=.5\pgf@y
		\advance\pgf@y by \pgf@ya
		\pgf@y=.5\pgf@y
	}
	\anchor{east north east}{
		\pgf@process{\southwest}
		\pgf@ya=.5\pgf@y
		\pgf@process{\northeast}
		\pgf@y=1.5\pgf@y
		\advance\pgf@y by \pgf@ya
		\pgf@y=.5\pgf@y
	}
	\anchor{east south east}{
		\pgf@process{\southwest}
		\pgf@ya=1.5\pgf@y
		\pgf@process{\northeast}
		\pgf@y=.5\pgf@y
		\advance\pgf@y by \pgf@ya
		\pgf@y=.5\pgf@y
	}
	\anchor{north north west}{
		\pgf@process{\southwest}
		\pgf@xa=1.5\pgf@x
		\pgf@process{\northeast}
		\pgf@x=.5\pgf@x
		\advance\pgf@x by \pgf@xa
		\pgf@x=.5\pgf@x
	}
	\anchor{north north east}{
		\pgf@process{\southwest}
		\pgf@xa=.5\pgf@x
		\pgf@process{\northeast}
		\pgf@x=1.5\pgf@x
		\advance\pgf@x by \pgf@xa
		\pgf@x=.5\pgf@x
	}
	\anchor{south south west}{
		\pgf@process{\northeast}
		\pgf@xa=.5\pgf@x
		\pgf@process{\southwest}
		\pgf@x=1.5\pgf@x
		\advance\pgf@x by \pgf@xa
		\pgf@x=.5\pgf@x
	}
	\anchor{south south east}{
		\pgf@process{\northeast}
		\pgf@xa=1.5\pgf@x
		\pgf@process{\southwest}
		\pgf@x=.5\pgf@x
		\advance\pgf@x by \pgf@xa
		\pgf@x=.5\pgf@x
	}
}

\pgfaddtoshape{basic}{
	\anchorlet{c}{center}		
	\anchorlet{n}{north}
	\anchorlet{e}{east}
	\anchorlet{s}{south}
	\anchorlet{w}{west}			
	\anchorlet{se}{south east}
	\anchorlet{sw}{south west}
	\anchorlet{ne}{north east}
	\anchorlet{nw}{north west}
	\anchorlet{wsw}{west south west}
	\anchorlet{wnw}{west north west}
	\anchorlet{ene}{east north east}
	\anchorlet{ese}{east south east}
	\anchorlet{nnw}{north north west}
	\anchorlet{nne}{north north east}
	\anchorlet{ssw}{south south west}
	\anchorlet{sse}{south south east}
}	

\pgfdeclareshape{ampshape}{
	\savedmacro\direction{
		\edef\direction{\pgfkeysvalueof{/tikz/ampkeys/direction}}%
	}
	\saveddimen\minwidth{
		\pgfmathsetlength\pgf@x{\pgfshapeminwidth}%
	}
	\saveddimen\minheight{
		\pgfmathsetlength\pgf@x{\pgfshapeminheight}%
	}
	\inheritsavedanchors[from=basic] 
	
	\inheritanchor[from=basic]{center}
	\inheritanchor[from=basic]{mid}		
	\inheritanchor[from=basic]{base}	
	\inheritanchor[from=basic]{north}
	\inheritanchor[from=basic]{south}		
	\inheritanchor[from=basic]{west}		
	\inheritanchor[from=basic]{mid west}				
	\inheritanchor[from=basic]{base west}		
	\inheritanchor[from=basic]{north west}		
	\inheritanchor[from=basic]{south west}		
	\inheritanchor[from=basic]{east}
	\inheritanchor[from=basic]{mid east}
	\inheritanchor[from=basic]{base east}	
	\inheritanchor[from=basic]{north east}
	\inheritanchor[from=basic]{south east}
	\inheritanchor[from=basic]{west south west}%
	\inheritanchor[from=basic]{west north west}%
	\inheritanchor[from=basic]{east north east}%
	\inheritanchor[from=basic]{east south east}%
	\inheritanchor[from=basic]{north north west}%
	\inheritanchor[from=basic]{north north east}%
	\inheritanchor[from=basic]{south south west}%
	\inheritanchor[from=basic]{south south east}%
	
	\inheritanchorborder[from=basic]	
	\inheritbackgroundpath[from=basic]

    \pgfutil@g@addto@macro\pgf@sh@s@ampshape{%
        \pgfutil@ifundefined{pgf@anchor@ampshape@in0}{
	        \expandafter\xdef\csname pgf@anchor@ampshape@in0\endcsname{%
	            \noexpand\ampshape@port{0}
	        }%
	    }{}%
        \pgfutil@ifundefined{pgf@anchor@ampshape@in}{
	        \expandafter\xdef\csname pgf@anchor@ampshape@in\endcsname{%
	            \noexpand\ampshape@port{0}
	        }%
	    }{}%
        \pgfutil@ifundefined{pgf@anchor@ampshape@out0}{
	        \expandafter\xdef\csname pgf@anchor@ampshape@out0\endcsname{%
	            \noexpand\ampshape@port{1}
	        }%
	    }{}%
        \pgfutil@ifundefined{pgf@anchor@ampshape@out}{
	        \expandafter\xdef\csname pgf@anchor@ampshape@out\endcsname{%
	            \noexpand\ampshape@port{1}
	        }%
	    }{}%
	}
}

\def\ampshape@port#1{
    \northeast	

    \ifnum#1=0	
	    \if\direction\direce
			\pgf@x=-\pgf@x
		    \pgf@ya= \pgf@y
		    \pgfmathsetlength{\pgf@y}{\pgf@ya-0.5*\minheight}%
		\fi
	    \if\direction\direcw
			\pgf@x=\pgf@x
		    \pgf@ya= \pgf@y
		    \pgfmathsetlength{\pgf@y}{\pgf@ya-0.5*\minheight}%
		\fi
	    \if\direction\direcn
			\pgf@y=-\pgf@y
		    \pgf@xa=\pgf@x
		    \pgfmathsetlength{\pgf@x}{\pgf@xa-0.5*\minwidth}%
		\fi
	    \if\direction\direcs
			\pgf@y=\pgf@y
		    \pgf@xa= \pgf@x
		    \pgfmathsetlength{\pgf@x}{\pgf@xa-0.5*\minwidth}%
		\fi
	\else	
	    \if\direction\direce
			\pgf@x=\pgf@x
		    \pgf@ya= \pgf@y
		    \pgfmathsetlength{\pgf@y}{\pgf@ya-0.5*\minheight}%
		\fi
	    \if\direction\direcw
			\pgf@x=-\pgf@x
		    \pgf@ya= \pgf@y
		    \pgfmathsetlength{\pgf@y}{\pgf@ya-0.5*\minheight}%
		\fi
	    \if\direction\direcn
			\pgf@y=\pgf@y
		    \pgf@xa= \pgf@x
		    \pgfmathsetlength{\pgf@x}{\pgf@xa-0.5*\minwidth}%
		\fi
	    \if\direction\direcs
			\pgf@y=-\pgf@y
		    \pgf@xa= \pgf@x
		    \pgfmathsetlength{\pgf@x}{\pgf@xa-0.5*\minwidth}%
		\fi
	\fi
}

\pgfaddtoshape{ampshape}{
	\anchorlet{c}{center}		
	\anchorlet{n}{north}
	\anchorlet{e}{east}
	\anchorlet{s}{south}
	\anchorlet{w}{west}			
	\anchorlet{se}{south east}
	\anchorlet{sw}{south west}
	\anchorlet{ne}{north east}
	\anchorlet{nw}{north west}
	\anchorlet{wsw}{west south west}
	\anchorlet{wnw}{west north west}
	\anchorlet{ene}{east north east}
	\anchorlet{ese}{east south east}
	\anchorlet{nnw}{north north west}
	\anchorlet{nne}{north north east}
	\anchorlet{ssw}{south south west}
	\anchorlet{sse}{south south east}
}

\tikzset{
	/tikz/ampkeys/.cd,
	height/.initial=0.5,
	width/.initial=0.5,
	color/.initial=O,
	direction/.initial=e,
	linestyle/.initial={linestyle, rounded corners = 0},
	/tikz/amp/.code={
		\pgfqkeys{/tikz/ampkeys}{#1}%
		\tikzset{/tikz/ampkeys/drawer/.expanded=%
			{\pgfkeysvalueof{/tikz/ampkeys/direction}}%
			{\pgfkeysvalueof{/tikz/ampkeys/height}}%
			{\pgfkeysvalueof{/tikz/ampkeys/width}}%
			{\pgfkeysvalueof{/tikz/ampkeys/color}}%
			{\pgfkeysvalueof{/tikz/ampkeys/linestyle}}%
		}
	},
	/tikz/ampkeys/drawer/.code n args={5}{%
		\tikzset{
			ampshape,
			minimum height=#2*\NODESIZE,
			minimum width=#3*\NODESIZE,
			append after command={
				\pgfextra{\let\bdr=\tikzlastnode%
				\if#1e
					\draw[draw=#4, fill=#4l, #5] (\bdr.sw) to (\bdr.nw) to (\bdr.e) to cycle {};
				\fi
				\if#1w
					\draw[draw=#4, fill=#4l, #5] (\bdr.se) to (\bdr.ne) to (\bdr.w) to cycle {};
				\fi
				\if#1n
					\draw[draw=#4, fill=#4l, #5] (\bdr.se) to (\bdr.sw) to (\bdr.n) to cycle {};
				\fi
				\if#1s
					\draw[draw=#4, fill=#4l, #5] (\bdr.ne) to (\bdr.nw) to (\bdr.s) to cycle {};
				\fi
				}
			}
		}
	},
}
\newcount\portcount

\pgfdeclareshape{aomshape}{
	\savedmacro\direction{
		\edef\direction{\pgfkeysvalueof{/tikz/aomkeys/direction}}%
	}
	\saveddimen\minwidth{
		\pgfmathsetlength\pgf@x{\pgfshapeminwidth}%
	}
	\saveddimen\minheight{
		\pgfmathsetlength\pgf@x{\pgfshapeminheight}%
	}
	\inheritsavedanchors[from=basic]

	\inheritanchor[from=basic]{center}
	\inheritanchor[from=basic]{mid}
	\inheritanchor[from=basic]{base}
	\inheritanchor[from=basic]{north}
	\inheritanchor[from=basic]{south}
	\inheritanchor[from=basic]{west}
	\inheritanchor[from=basic]{mid west}
	\inheritanchor[from=basic]{base west}
	\inheritanchor[from=basic]{north west}
	\inheritanchor[from=basic]{south west}
	\inheritanchor[from=basic]{east}
	\inheritanchor[from=basic]{mid east}
	\inheritanchor[from=basic]{base east}
	\inheritanchor[from=basic]{north east}
	\inheritanchor[from=basic]{south east}
	\inheritanchor[from=basic]{west south west}%
	\inheritanchor[from=basic]{west north west}%
	\inheritanchor[from=basic]{east north east}%
	\inheritanchor[from=basic]{east south east}%
	\inheritanchor[from=basic]{north north west}%
	\inheritanchor[from=basic]{north north east}%
	\inheritanchor[from=basic]{south south west}%
	\inheritanchor[from=basic]{south south east}%

	\inheritanchorborder[from=basic]
	\inheritbackgroundpath[from=basic]

	\pgfutil@g@addto@macro\pgf@sh@s@aomshape{%
		\pgfutil@ifundefined{pgf@anchor@aomshape@in0}{
			\expandafter\xdef\csname pgf@anchor@aomshape@in0\endcsname{%
				\noexpand\aomshape@port{0}
			}%
		}{}%
		\pgfutil@ifundefined{pgf@anchor@aomshape@in}{
			\expandafter\xdef\csname pgf@anchor@aomshape@in\endcsname{%
				\noexpand\aomshape@port{0}
			}%
		}{}%
		\pgfutil@ifundefined{pgf@anchor@aomshape@out0}{
			\expandafter\xdef\csname pgf@anchor@aomshape@out0\endcsname{%
				\noexpand\aomshape@port{1}
			}%
		}{}%
		\pgfutil@ifundefined{pgf@anchor@aomshape@out}{
			\expandafter\xdef\csname pgf@anchor@aomshape@out\endcsname{%
				\noexpand\aomshape@port{1}
			}%
		}{}%
	}
}

\def\aomshape@port#1{
	\northeast	

	\ifnum#1=0	
		\if\direction\direce
			\pgf@x=-\pgf@x
			\pgf@ya= \pgf@y
			\pgfmathsetlength{\pgf@y}{\pgf@ya-0.5*\minheight}%
		\fi
		\if\direction\direcw
			\pgf@x=\pgf@x
			\pgf@ya= \pgf@y
			\pgfmathsetlength{\pgf@y}{\pgf@ya-0.5*\minheight}%
		\fi
		\if\direction\direcn
			\pgf@y=-\pgf@y
			\pgf@xa=\pgf@x
			\pgfmathsetlength{\pgf@x}{\pgf@xa-0.5*\minwidth}%
		\fi
		\if\direction\direcs
			\pgf@y=\pgf@y
			\pgf@xa= \pgf@x
			\pgfmathsetlength{\pgf@x}{\pgf@xa-0.5*\minwidth}%
		\fi
	\else	
		\if\direction\direce
			\pgf@x=\pgf@x
			\pgf@ya= \pgf@y
			\pgfmathsetlength{\pgf@y}{\pgf@ya-0.5*\minheight}%
		\fi
		\if\direction\direcw
			\pgf@x=-\pgf@x
			\pgf@ya= \pgf@y
			\pgfmathsetlength{\pgf@y}{\pgf@ya-0.5*\minheight}%
		\fi
		\if\direction\direcn
			\pgf@y=\pgf@y
			\pgf@xa= \pgf@x
			\pgfmathsetlength{\pgf@x}{\pgf@xa-0.5*\minwidth}%
		\fi
		\if\direction\direcs
			\pgf@y=-\pgf@y
			\pgf@xa= \pgf@x
			\pgfmathsetlength{\pgf@x}{\pgf@xa-0.5*\minwidth}%
		\fi
	\fi
}

\pgfaddtoshape{aomshape}{
	\anchorlet{c}{center}
	\anchorlet{n}{north}
	\anchorlet{e}{east}
	\anchorlet{s}{south}
	\anchorlet{w}{west}
	\anchorlet{se}{south east}
	\anchorlet{sw}{south west}
	\anchorlet{ne}{north east}
	\anchorlet{nw}{north west}
	\anchorlet{wsw}{west south west}
	\anchorlet{wnw}{west north west}
	\anchorlet{ene}{east north east}
	\anchorlet{ese}{east south east}
	\anchorlet{nnw}{north north west}
	\anchorlet{nne}{north north east}
	\anchorlet{ssw}{south south west}
	\anchorlet{sse}{south south east}
}

\tikzset{
/tikz/aomkeys/.cd,
size/.initial=1,
circlesize/.initial=1,
color/.initial=O,
direction/.initial=e,
linestyle/.initial={linestyle, inner sep=0.5mm},
fillgradient/.initial=O,
/tikz/aom/.code={
\pgfqkeys{/tikz/aomkeys}{#1}%
\tikzset{/tikz/aomkeys/drawer/.expanded=%
	{\pgfkeysvalueof{/tikz/aomkeys/size}}%
	{\pgfkeysvalueof{/tikz/aomkeys/color}}%
	{\pgfkeysvalueof{/tikz/aomkeys/linestyle}}%
\if\pgfkeysvalueof{/tikz/aomkeys/direction}e
	{0}%
\fi
\if\pgfkeysvalueof{/tikz/aomkeys/direction}w
	{0}%
\fi
\if\pgfkeysvalueof{/tikz/aomkeys/direction}n
	{1}%
\fi
\if\pgfkeysvalueof{/tikz/aomkeys/direction}s
	{1}%
\fi
{\pgfkeysvalueof{/tikz/aomkeys/direction}}%
{\pgfkeysvalueof{/tikz/aomkeys/circlesize}}%
{\pgfkeysvalueof{/tikz/aomkeys/fillgradient}}%
}
},
/tikz/aomkeys/drawer/.code n args={7}{%
		\tikzset{
			aomshape,
			draw,
			minimum height = #1*\NODESIZE,
			minimum width = #1*\NODESIZE,
			#2,
			#3,
			append after command={
					\pgfextra{\let\bdr=\tikzlastnode%
						\node[#7, fit=(\bdr.nw)(\bdr.se)] (boxgradient){};

						\node[coordinate] at ($(\bdr.in)!0.25!(\bdr.out)$) (circlein){};
						\node[coordinate] at ($(\bdr.in)!0.75!(\bdr.out)$) (circleout){};

						\ifnum#4>0
							\node[coordinate] at (circleout -| \bdr.nne) (circleouttop){};
						\else
							\node[coordinate] at (circleout |- \bdr.ene) (circleouttop){};
						\fi

						\draw[---, #2, #3, fill] (\bdr.in) to (circlein) circle (0.05*#6);
						\draw[---, #2, #3, fill] (\bdr.out) to (circleout) circle (0.05*#6);

						\draw[---, #2, #3] (circlein) to (circleouttop){};

					}
				}
		}
	},
}
\newcount\portcount

\pgfdeclareshape{boxshape}{
	\savedmacro\nin{
		\edef\nin{\pgfkeysvalueof{/tikz/boxkeys/nin}}%
	}
	\savedmacro\nout{
		\edef\nout{\pgfkeysvalueof{/tikz/boxkeys/nout}}%
	}
	\savedmacro\direction{
		\edef\direction{\pgfkeysvalueof{/tikz/boxkeys/direction}}%
	}
	\inheritsavedanchors[from=basic]

	\inheritanchor[from=basic]{center}
	\inheritanchor[from=basic]{mid}
	\inheritanchor[from=basic]{base}
	\inheritanchor[from=basic]{north}
	\inheritanchor[from=basic]{south}
	\inheritanchor[from=basic]{west}
	\inheritanchor[from=basic]{mid west}
	\inheritanchor[from=basic]{base west}
	\inheritanchor[from=basic]{north west}
	\inheritanchor[from=basic]{south west}
	\inheritanchor[from=basic]{east}
	\inheritanchor[from=basic]{mid east}
	\inheritanchor[from=basic]{base east}
	\inheritanchor[from=basic]{north east}
	\inheritanchor[from=basic]{south east}
	\inheritanchor[from=basic]{west south west}%
	\inheritanchor[from=basic]{west north west}%
	\inheritanchor[from=basic]{east north east}%
	\inheritanchor[from=basic]{east south east}%
	\inheritanchor[from=basic]{north north west}%
	\inheritanchor[from=basic]{north north east}%
	\inheritanchor[from=basic]{south south west}%
	\inheritanchor[from=basic]{south south east}%

	\inheritanchorborder[from=basic]
	\inheritbackgroundpath[from=basic]

	\pgfutil@g@addto@macro\pgf@sh@s@boxshape{%
		\pgfmathsetcount{\portcount}{0}
		\pgfmathloop%
		\ifnum\the\portcount<\nin
		\pgfutil@ifundefined{pgf@anchor@boxshape@in\the\portcount}{
			\expandafter\xdef\csname pgf@anchor@boxshape@in\the\portcount\endcsname{%
				\noexpand\boxshape@port[\the\portcount]{0}
			}%
		}{}%
		\ifnum\the\portcount=0
			\pgfutil@ifundefined{pgf@anchor@boxshape@in}{%
				\expandafter\xdef\csname pgf@anchor@boxshape@in\endcsname{%
					\noexpand\boxshape@port[\the\portcount]{0}
				}%
			}{}%
		\fi
		\pgfmathaddtocount{\portcount}{1}	
		\repeatpgfmathloop
		\pgfmathsetcount{\portcount}{0}
		\pgfmathloop%
		\ifnum\the\portcount<\nout
		\pgfutil@ifundefined{pgf@anchor@boxshape@out\the\portcount}{%
			\expandafter\xdef\csname pgf@anchor@boxshape@out\the\portcount\endcsname{%
				\noexpand\boxshape@port[\the\portcount]{1}
			}%
		}{}%
		\ifnum\the\portcount=0
			\pgfutil@ifundefined{pgf@anchor@boxshape@out}{%
				\expandafter\xdef\csname pgf@anchor@boxshape@out\endcsname{%
					\noexpand\boxshape@port[\the\portcount]{1}
				}%
			}{}%
		\fi
		\pgfmathaddtocount{\portcount}{1}	
		\repeatpgfmathloop
	}
}

\def\boxshape@port[#1]#2{
	\northeast \pgf@xa=\pgf@x \pgf@ya=\pgf@y
	\southwest \pgf@xb=\pgf@x \pgf@yb=\pgf@y

	\ifnum#2=0	
		\if\direction\direce	
			\pgf@x=\pgf@xb
			\pgf@yc=\pgf@ya \advance\pgf@yc by -\pgf@yb	
			\pgfmathsetlength{\pgf@y}{\pgf@ya-(#1 + 0.5)*(\pgf@yc/\nin)}%
		\fi
		\if\direction\direcw
			\pgf@x=\pgf@xa
			\pgf@yc=\pgf@ya \advance\pgf@yc by -\pgf@yb	
			\pgfmathsetlength{\pgf@y}{\pgf@ya-(#1 + 0.5)*(\pgf@yc/\nin)}%
		\fi
		\if\direction\direcn
			\pgf@y=\pgf@yb
			\pgf@xc=\pgf@xa \advance\pgf@xc by -\pgf@xb	
			\pgfmathsetlength{\pgf@x}{\pgf@xb+(#1 + 0.5)*(\pgf@xc/\nin)}%
		\fi
		\if\direction\direcs
			\pgf@y=\pgf@ya
			\pgf@xc=\pgf@xa \advance\pgf@xc by -\pgf@xb	
			\pgfmathsetlength{\pgf@x}{\pgf@xb+(#1 + 0.5)*(\pgf@xc/\nin)}%
		\fi
	\else	
		\if\direction\direce	
			\pgf@x=\pgf@xa
			\pgf@yc=\pgf@ya \advance\pgf@yc by -\pgf@yb	
			\pgfmathsetlength{\pgf@y}{\pgf@ya-(#1 + 0.5)*(\pgf@yc/\nout)}%
		\fi
		\if\direction\direcw
			\pgf@x=\pgf@xb
			\pgf@yc=\pgf@ya \advance\pgf@yc by -\pgf@yb	
			\pgfmathsetlength{\pgf@y}{\pgf@ya-(#1 + 0.5)*(\pgf@yc/\nout)}%
		\fi
		\if\direction\direcn
			\pgf@y=\pgf@ya
			\pgf@xc=\pgf@xa \advance\pgf@xc by -\pgf@xb	
			\pgfmathsetlength{\pgf@x}{\pgf@xb+(#1 + 0.5)*(\pgf@xc/\nout)}%
		\fi
		\if\direction\direcs
			\pgf@y=\pgf@yb
			\pgf@xc=\pgf@xa \advance\pgf@xc by -\pgf@xb	
			\pgfmathsetlength{\pgf@x}{\pgf@xb+(#1 + 0.5)*(\pgf@xc/\nout)}%
		\fi
	\fi
}

\pgfaddtoshape{boxshape}{
	\anchorlet{c}{center}
	\anchorlet{n}{north}
	\anchorlet{e}{east}
	\anchorlet{s}{south}
	\anchorlet{w}{west}
	\anchorlet{se}{south east}
	\anchorlet{sw}{south west}
	\anchorlet{ne}{north east}
	\anchorlet{nw}{north west}
	\anchorlet{wsw}{west south west}
	\anchorlet{wnw}{west north west}
	\anchorlet{ene}{east north east}
	\anchorlet{ese}{east south east}
	\anchorlet{nnw}{north north west}
	\anchorlet{nne}{north north east}
	\anchorlet{ssw}{south south west}
	\anchorlet{sse}{south south east}
}

\tikzset{
/tikz/boxkeys/.cd,
height/.initial=0.5,
width/.initial=1,
color/.initial=O,
direction/.initial=e,
linestyle/.initial={linestyle, inner sep=0.5mm},
nin/.initial=1,
nout/.initial=1,
draw/.initial=1,
/tikz/box/.code={
\pgfqkeys{/tikz/boxkeys}{#1}%
\tikzset{/tikz/boxkeys/drawer/.expanded=%
\if\pgfkeysvalueof{/tikz/boxkeys/direction}e
	{\pgfkeysvalueof{/tikz/boxkeys/width}}%
	{\pgfkeysvalueof{/tikz/boxkeys/height}}%
	{0}%
	{-90}%
\fi
\if\pgfkeysvalueof{/tikz/boxkeys/direction}w
	{\pgfkeysvalueof{/tikz/boxkeys/width}}%
	{\pgfkeysvalueof{/tikz/boxkeys/height}}%
	{0}%
	{90}%
\fi
\if\pgfkeysvalueof{/tikz/boxkeys/direction}n
	{\pgfkeysvalueof{/tikz/boxkeys/width}}%
	{\pgfkeysvalueof{/tikz/boxkeys/height}}%
	{1}%
	{0}%
\fi
\if\pgfkeysvalueof{/tikz/boxkeys/direction}s
	{\pgfkeysvalueof{/tikz/boxkeys/width}}%
	{\pgfkeysvalueof{/tikz/boxkeys/height}}%
	{1}%
	{180}%
\fi
{\pgfkeysvalueof{/tikz/boxkeys/color}}%
{\pgfkeysvalueof{/tikz/boxkeys/linestyle}}%
\ifnum\pgfkeysvalueof{/tikz/boxkeys/draw}>0%
	{draw}%
\else
	{}
\fi
}
},
/tikz/boxkeys/drawer/.code n args={7}{%
		\tikzset{
			boxshape,
			#7,
			#6,
			#5,
			minimum height=
			\ifnum#3>0	
				#1*\NODESIZE
			\else
				#2*\NODESIZE
			\fi
			,minimum width=
			\ifnum#3>0
				#2*\NODESIZE
			\else
				#1*\NODESIZE
			\fi
		}
	},
}
\pgfdeclareshape{beshape}{ 
    \savedmacro\nports{
        \edef\nports{\pgfkeysvalueof{/tikz/bekeys/nports}}%
    }
    \savedmacro\direction{
        \edef\direction{\pgfkeysvalueof{/tikz/bekeys/direction}}%
    }
    \savedmacro\inverted{
        \edef\inverted{\pgfkeysvalueof{/tikz/bekeys/inverted}}%
    }
    \savedmacro\ninports{
        \ifnum\inverted=0
        \edef\ninports{\nports}
        \else
        \edef\ninports{1}%
        \fi
    }
    \savedmacro\noutports{
        \ifnum\inverted=0
        \edef\noutports{1}%
        \else
        \edef\noutports{\nports}%
        \fi
    }
    \inheritsavedanchors[from=basic]

    \inheritanchor[from=basic]{center}
    \inheritanchor[from=basic]{mid}
    \inheritanchor[from=basic]{base}
    \inheritanchor[from=basic]{north}
    \inheritanchor[from=basic]{south}
    \inheritanchor[from=basic]{west}
    \inheritanchor[from=basic]{mid west}
    \inheritanchor[from=basic]{base west}
    \inheritanchor[from=basic]{north west}
    \inheritanchor[from=basic]{south west}
    \inheritanchor[from=basic]{east}
    \inheritanchor[from=basic]{mid east}
    \inheritanchor[from=basic]{base east}
    \inheritanchor[from=basic]{north east}
    \inheritanchor[from=basic]{south east}
    \inheritanchor[from=basic]{west south west}%
    \inheritanchor[from=basic]{west north west}%
    \inheritanchor[from=basic]{east north east}%
    \inheritanchor[from=basic]{east south east}%
    \inheritanchor[from=basic]{north north west}%
    \inheritanchor[from=basic]{north north east}%
    \inheritanchor[from=basic]{south south west}%
    \inheritanchor[from=basic]{south south east}%

    \inheritanchorborder[from=basic]
    \inheritbackgroundpath[from=basic]

    \pgfutil@g@addto@macro\pgf@sh@s@beshape{%
        \pgfmathsetcount{\portcount}{0}
        \pgfmathloop%
        \ifnum\the\portcount<\nports
        \ifnum\the\portcount<\ninports
        \pgfutil@ifundefined{pgf@anchor@beshape@in\the\portcount}{
            \expandafter\xdef\csname pgf@anchor@beshape@in\the\portcount\endcsname{%
                \noexpand\beshape@port[\the\portcount]{0}
            }%
        }{}%
        \ifnum\the\portcount=0
        \pgfutil@ifundefined{pgf@anchor@beshape@in}{%
            \expandafter\xdef\csname pgf@anchor@beshape@in\endcsname{%
                \noexpand\beshape@port[\the\portcount]{0}
            }%
        }{}%
        \fi
        \fi
        \ifnum\the\portcount<\noutports
        \pgfutil@ifundefined{pgf@anchor@beshape@out\the\portcount}{%
            \expandafter\xdef\csname pgf@anchor@beshape@out\the\portcount\endcsname{%
                \noexpand\beshape@port[\the\portcount]{1}
            }%
        }{}%
        \ifnum\the\portcount=0
        \pgfutil@ifundefined{pgf@anchor@beshape@out}{%
            \expandafter\xdef\csname pgf@anchor@beshape@out\endcsname{%
                \noexpand\beshape@port[\the\portcount]{1}
            }%
        }{}%
        \fi
        \fi
        \pgfmathaddtocount{\portcount}{1}    
        \repeatpgfmathloop
    }
}

\def\beshape@port[#1]#2{
    \northeast \pgf@xa=\pgf@x \pgf@ya=\pgf@y
    \southwest \pgf@xb=\pgf@x \pgf@yb=\pgf@y

    \ifnum#2=0
    \ifnum\inverted=0
    \def\chooseports{0}
    \else
    \def\chooseports{1}
    \fi
    \else
    \ifnum\inverted=0
    \def\chooseports{1}
    \else
    \def\chooseports{0}
    \fi
    \fi

    \ifnum\chooseports=0    
    \if\direction\direce
    \pgf@x=\pgf@xb
    \pgf@yc=\pgf@ya \advance\pgf@yc by -\pgf@yb    
    \pgfmathsetlength{\pgf@y}{\pgf@ya-(#1 + 0.5)*(\pgf@yc/\nports)}%
    \fi
    \if\direction\direcw
    \pgf@x=\pgf@xa
    \pgf@yc=\pgf@ya \advance\pgf@yc by -\pgf@yb    
    \pgfmathsetlength{\pgf@y}{\pgf@ya-(#1 + 0.5)*(\pgf@yc/\nports)}%
    \fi
    \if\direction\direcn
    \pgf@y=\pgf@yb
    \pgf@xc=\pgf@xa \advance\pgf@xc by -\pgf@xb    
    \pgfmathsetlength{\pgf@x}{\pgf@xb+(#1 + 0.5)*(\pgf@xc/\nports)}%
    \fi
    \if\direction\direcs
    \pgf@y=\pgf@ya
    \pgf@xc=\pgf@xa \advance\pgf@xc by -\pgf@xb    
    \pgfmathsetlength{\pgf@x}{\pgf@xb+(#1 + 0.5)*(\pgf@xc/\nports)}%
    \fi
    \else    
    \if\direction\direce
    \pgf@x=\pgf@xa
    \pgf@yc=\pgf@ya \advance\pgf@yc by -\pgf@yb    
    \pgfmathsetlength{\pgf@y}{\pgf@ya-0.5\pgf@yc}%
    \fi
    \if\direction\direcw
    \pgf@x=\pgf@xb
    \pgf@yc=\pgf@ya \advance\pgf@yc by -\pgf@yb    
    \pgfmathsetlength{\pgf@y}{\pgf@ya-0.5\pgf@yc}%
    \fi
    \if\direction\direcn
    \pgf@y=\pgf@ya
    \pgf@xc=\pgf@xa \advance\pgf@xc by -\pgf@xb    
    \pgfmathsetlength{\pgf@x}{\pgf@xa-0.5\pgf@xc}%
    \fi
    \if\direction\direcs
    \pgf@y=\pgf@yb
    \pgf@xc=\pgf@xa \advance\pgf@xc by -\pgf@xb    
    \pgfmathsetlength{\pgf@x}{\pgf@xa-0.5\pgf@xc}%
    \fi
    \fi
}

\pgfaddtoshape{beshape}{
    \anchorlet{c}{center}
    \anchorlet{n}{north}
    \anchorlet{e}{east}
    \anchorlet{s}{south}
    \anchorlet{w}{west}
    \anchorlet{se}{south east}
    \anchorlet{sw}{south west}
    \anchorlet{ne}{north east}
    \anchorlet{nw}{north west}
    \anchorlet{wsw}{west south west}
    \anchorlet{wnw}{west north west}
    \anchorlet{ene}{east north east}
    \anchorlet{ese}{east south east}
    \anchorlet{nnw}{north north west}
    \anchorlet{nne}{north north east}
    \anchorlet{ssw}{south south west}
    \anchorlet{sse}{south south east}
}

\tikzset{
    /tikz/bekeys/.cd,
    height/.initial=1,
    width/.initial=0.5,
    color/.initial=O,
    direction/.initial=e,
    linestyle/.initial={linestyle, rounded corners = 0},
    nports/.initial=3,
    inverted/.initial=0,    
    sbe/.initial=0,
    angle/.initial=60,
    /tikz/be/.code={
        \pgfqkeys{/tikz/bekeys}{#1}%
        \tikzset{/tikz/bekeys/drawer/.expanded=%
            \if\pgfkeysvalueof{/tikz/bekeys/direction}e
                {\pgfkeysvalueof{/tikz/bekeys/width}}%
                {\pgfkeysvalueof{/tikz/bekeys/height}}%
                {0}%
                {-90}%
            \fi
            \if\pgfkeysvalueof{/tikz/bekeys/direction}w
                {\pgfkeysvalueof{/tikz/bekeys/width}}%
                {\pgfkeysvalueof{/tikz/bekeys/height}}%
                {0}%
                {90}%
            \fi
            \if\pgfkeysvalueof{/tikz/bekeys/direction}n
                {\pgfkeysvalueof{/tikz/bekeys/width}}%
                {\pgfkeysvalueof{/tikz/bekeys/height}}%
                {1}%
                {0}%
            \fi
            \if\pgfkeysvalueof{/tikz/bekeys/direction}s
                {\pgfkeysvalueof{/tikz/bekeys/width}}%
                {\pgfkeysvalueof{/tikz/bekeys/height}}%
                {1}%
                {180}%
            \fi
            {\pgfkeysvalueof{/tikz/bekeys/color}}%
            {\pgfkeysvalueof{/tikz/bekeys/linestyle}}%
            {\pgfkeysvalueof{/tikz/bekeys/sbe}}%
            {\pgfkeysvalueof{/tikz/bekeys/angle}}%
        }
    },
    /tikz/bekeys/drawer/.code n args={8}{%
        \tikzset{
            beshape,
            #6,
            #5,
            minimum height=
            \ifnum#3>0    
            #1*\NODESIZE
            \else
            #2*\NODESIZE
            \fi
            ,minimum width=
            \ifnum#3>0
            #2*\NODESIZE
            \else
            #1*\NODESIZE
            \fi
            ,append after command={
                \pgfextra{
                    \let\bdr=\tikzlastnode%
                    \node[trapezium, line width = \NODETHICKNESS, minimum height=#1*\NODESIZE, minimum width=#2*\NODESIZE, trapezium stretches=true, rotate=#4, trapezium angle=70, inner sep=0.001mm, #5, #6] at (\bdr) (trap) {};

                    \node[rectangle, line width = \NODETHICKNESS, minimum height=#1*\NODESIZE, minimum width=#2*\NODESIZE, anchor=north, rotate=#4, #5, #6] at (trap.south) (r1) {};

                    \tikzmath{coordinate \C;
                    \C = (trap.top left corner)-(trap.top right corner);
                    \distAB = sqrt((\Cx)^2+(\Cy)^2);
                    }

                    \node[rectangle, line width = \NODETHICKNESS, minimum height=#1*\NODESIZE, minimum width=\distAB, anchor=south, rotate=#4, #5, #6, red] at (trap.north) (r2) {};

					\draw[#5, #6] (trap.top left corner) to (r2.north west) to (r2.north east) to (trap.top right corner) to (trap.bottom right corner) to (r1.south east) to (r1.south west) to (r1.north west) to cycle;

                }
            }
        }
    },
}
\pgfdeclareshape{couplershape}{	

	\inheritsavedanchors[from=basic] 

	\inheritanchor[from=basic]{center}
	\inheritanchor[from=basic]{mid}		
	\inheritanchor[from=basic]{base}	
	\inheritanchor[from=basic]{north}
	\inheritanchor[from=basic]{south}		
	\inheritanchor[from=basic]{west}		
	\inheritanchor[from=basic]{mid west}				
	\inheritanchor[from=basic]{base west}		
	\inheritanchor[from=basic]{north west}		
	\inheritanchor[from=basic]{south west}		
	\inheritanchor[from=basic]{east}
	\inheritanchor[from=basic]{mid east}
	\inheritanchor[from=basic]{base east}	
	\inheritanchor[from=basic]{north east}
	\inheritanchor[from=basic]{south east}
	\inheritanchor[from=basic]{west south west}%
	\inheritanchor[from=basic]{west north west}%
	\inheritanchor[from=basic]{east north east}%
	\inheritanchor[from=basic]{east south east}%
	\inheritanchor[from=basic]{north north west}%
	\inheritanchor[from=basic]{north north east}%
	\inheritanchor[from=basic]{south south west}%
	\inheritanchor[from=basic]{south south east}%
	
	\inheritanchorborder[from=basic]	
	\inheritbackgroundpath[from=basic]
}

\pgfaddtoshape{couplershape}{
	\anchorlet{in}{center}		
	\anchorlet{in0}{center}
	\anchorlet{out}{center}
	\anchorlet{out0}{center}
	\anchorlet{c}{center}		
	\anchorlet{n}{north}
	\anchorlet{e}{east}
	\anchorlet{s}{south}
	\anchorlet{w}{west}			
	\anchorlet{se}{south east}
	\anchorlet{sw}{south west}
	\anchorlet{ne}{north east}
	\anchorlet{nw}{north west}
	\anchorlet{wsw}{west south west}
	\anchorlet{wnw}{west north west}
	\anchorlet{ene}{east north east}
	\anchorlet{ese}{east south east}
	\anchorlet{nnw}{north north west}
	\anchorlet{nne}{north north east}
	\anchorlet{ssw}{south south west}
	\anchorlet{sse}{south south east}
}

\tikzset{
	/tikz/couplerkeys/.cd,
	size/.initial=0.2,
	color/.initial=O,
	rotation/.initial=0,
	heightwidthratio/.initial=0.5,
	/tikz/coupler/.code={
		\pgfqkeys{/tikz/couplerkeys}{#1}%
		\tikzset{/tikz/couplerkeys/drawer/.expanded=%
			{\pgfkeysvalueof{/tikz/couplerkeys/size}}%
			{\pgfkeysvalueof{/tikz/couplerkeys/color}}%
			{\pgfkeysvalueof{/tikz/couplerkeys/rotation}}%
			{\pgfkeysvalueof{/tikz/couplerkeys/heightwidthratio}}%
		}
	},
	/tikz/couplerkeys/drawer/.code n args={4}{%
		\tikzset{
			couplershape,
			minimum height=#1*\NODESIZE
			\ifnum#3<1
				\ifnum#3>-1
					*#4
				\fi
			\fi
			,minimum width=#1*\NODESIZE
			\ifnum#3<91
				\ifnum#3>89
					*#4
				\fi
			\fi
			\ifnum#3<-89
				\ifnum#3>-91
					*#4
				\fi
			\fi
			,#2,
			append after command={
				\pgfextra{\let\bdr=\tikzlastnode%
				\node[ellipse, fill, #2, rotate=#3, outer sep = 0, minimum width=#1*\NODESIZE, minimum height=#1*#4*\NODESIZE] at (\bdr.center){};
				}
			}
		}
	},
}
\pgfdeclareshape{fibershape}{
	\savedmacro\direction{
		\edef\direction{\pgfkeysvalueof{/tikz/fiberkeys/direction}}%
	}
	\savedmacro\flip{
		\edef\flip{\pgfkeysvalueof{/tikz/fiberkeys/flip}}%
	}
	\saveddimen\minwidth{
		\pgfmathsetlength\pgf@x{\pgfshapeminwidth}%
	}
	\saveddimen\minheight{
		\pgfmathsetlength\pgf@x{\pgfshapeminheight}%
	}
	\inheritsavedanchors[from=basic] 
	
	\inheritanchor[from=basic]{center}
	\inheritanchor[from=basic]{mid}		
	\inheritanchor[from=basic]{base}	
	\inheritanchor[from=basic]{north}
	\inheritanchor[from=basic]{south}		
	\inheritanchor[from=basic]{west}		
	\inheritanchor[from=basic]{mid west}				
	\inheritanchor[from=basic]{base west}		
	\inheritanchor[from=basic]{north west}		
	\inheritanchor[from=basic]{south west}		
	\inheritanchor[from=basic]{east}
	\inheritanchor[from=basic]{mid east}
	\inheritanchor[from=basic]{base east}	
	\inheritanchor[from=basic]{north east}
	\inheritanchor[from=basic]{south east}
	\inheritanchor[from=basic]{west south west}%
	\inheritanchor[from=basic]{west north west}%
	\inheritanchor[from=basic]{east north east}%
	\inheritanchor[from=basic]{east south east}%
	\inheritanchor[from=basic]{north north west}%
	\inheritanchor[from=basic]{north north east}%
	\inheritanchor[from=basic]{south south west}%
	\inheritanchor[from=basic]{south south east}%
	
	\inheritanchorborder[from=basic]	
	\inheritbackgroundpath[from=basic]

    \pgfutil@g@addto@macro\pgf@sh@s@fibershape{%
        \pgfutil@ifundefined{pgf@anchor@fibershape@in0}{
	        \expandafter\xdef\csname pgf@anchor@fibershape@in0\endcsname{%
	            \noexpand\fibershape@port{0}
	        }%
	    }{}%
        \pgfutil@ifundefined{pgf@anchor@fibershape@in}{
	        \expandafter\xdef\csname pgf@anchor@fibershape@in\endcsname{%
	            \noexpand\fibershape@port{0}
	        }%
	    }{}%
        \pgfutil@ifundefined{pgf@anchor@fibershape@out0}{
	        \expandafter\xdef\csname pgf@anchor@fibershape@out0\endcsname{%
	            \noexpand\fibershape@port{1}
	        }%
	    }{}%
        \pgfutil@ifundefined{pgf@anchor@fibershape@out}{
	        \expandafter\xdef\csname pgf@anchor@fibershape@out\endcsname{%
	            \noexpand\fibershape@port{1}
	        }%
	    }{}%
	}
}

\def\fibershape@port#1{
    \northeast	

    \ifnum#1=0	
	    \if\direction\direce
			\pgf@x=-\pgf@x
	    	\if\flip\flipfalse
		    	\pgf@y=-\pgf@y
		    \else
		    	\pgf@y=\pgf@y
		    \fi
		\fi
	    \if\direction\direcw
			\pgf@x=\pgf@x
	    	\if\flip\flipfalse
		    	\pgf@y=-\pgf@y
		    \else
		    	\pgf@y=\pgf@y
		    \fi
		\fi
	    \if\direction\direcn
			\pgf@y=-\pgf@y
	    	\if\flip\flipfalse
		    	\pgf@x=-\pgf@x
		    \else
		    	\pgf@x=\pgf@x
		    \fi
		\fi
	    \if\direction\direcs
			\pgf@y=\pgf@y
	    	\if\flip\flipfalse
		    	\pgf@x=-\pgf@x
		    \else
		    	\pgf@x=\pgf@x
		    \fi
		\fi
	\else	
	    \if\direction\direce
			\pgf@x=\pgf@x
	    	\if\flip\flipfalse
		    	\pgf@y=-\pgf@y
		    \else
		    	\pgf@y=\pgf@y
		    \fi
		\fi
	    \if\direction\direcw
			\pgf@x=-\pgf@x
	    	\if\flip\flipfalse
		    	\pgf@y=-\pgf@y
		    \else
		    	\pgf@y=\pgf@y
		    \fi
		\fi
	    \if\direction\direcn
			\pgf@y=\pgf@y
	    	\if\flip\flipfalse
		    	\pgf@x=-\pgf@x
		    \else
		    	\pgf@x=\pgf@x
		    \fi
		\fi
	    \if\direction\direcs
			\pgf@y=-\pgf@y
	    	\if\flip\flipfalse
		    	\pgf@x=-\pgf@x
		    \else
		    	\pgf@x=\pgf@x
		    \fi
		\fi
	\fi
}

\pgfaddtoshape{fibershape}{
	\anchorlet{c}{center}		
	\anchorlet{n}{north}
	\anchorlet{e}{east}
	\anchorlet{s}{south}
	\anchorlet{w}{west}			
	\anchorlet{se}{south east}
	\anchorlet{sw}{south west}
	\anchorlet{ne}{north east}
	\anchorlet{nw}{north west}
	\anchorlet{wsw}{west south west}
	\anchorlet{wnw}{west north west}
	\anchorlet{ene}{east north east}
	\anchorlet{ese}{east south east}
	\anchorlet{nnw}{north north west}
	\anchorlet{nne}{north north east}
	\anchorlet{ssw}{south south west}
	\anchorlet{sse}{south south east}
}

\tikzset{
	/tikz/fiberkeys/.cd,
	size/.initial=1,
	color/.initial=C0,
	direction/.initial=e,
	linestyle/.initial={linestyle},
	flip/.initial={0},
	drawbase/.initial={1},
	/tikz/fiber/.code={
		\pgfqkeys{/tikz/fiberkeys}{#1}%
		\tikzset{/tikz/fiberkeys/drawer/.expanded=%
			{\pgfkeysvalueof{/tikz/fiberkeys/direction}}%
			{\pgfkeysvalueof{/tikz/fiberkeys/size}}%
			{\pgfkeysvalueof{/tikz/fiberkeys/color}}%
			{\pgfkeysvalueof{/tikz/fiberkeys/linestyle}}%
			\if\pgfkeysvalueof{/tikz/fiberkeys/direction}e
				{a}%
				\if\pgfkeysvalueof{/tikz/fiberkeys/flip}0
					{south}%
				\else
					{north}%
				\fi
				{\pgfkeysvalueof{/tikz/fiberkeys/size}}
				{\pgfkeysvalueof{/tikz/fiberkeys/size} * 0.5}
			\fi
			\if\pgfkeysvalueof{/tikz/fiberkeys/direction}w
				{a}%
				\if\pgfkeysvalueof{/tikz/fiberkeys/flip}0
					{south}%
				\else
					{north}%
				\fi
				{\pgfkeysvalueof{/tikz/fiberkeys/size}}
				{\pgfkeysvalueof{/tikz/fiberkeys/size} * 0.5}
			\fi
			\if\pgfkeysvalueof{/tikz/fiberkeys/direction}n
				{b}%
				\if\pgfkeysvalueof{/tikz/fiberkeys/flip}0
					{west}%
				\else
					{east}%
				\fi
				{\pgfkeysvalueof{/tikz/fiberkeys/size} * 0.5}
				{\pgfkeysvalueof{/tikz/fiberkeys/size}}
			\fi
			\if\pgfkeysvalueof{/tikz/fiberkeys/direction}s
				{b}%
				\if\pgfkeysvalueof{/tikz/fiberkeys/flip}0
					{west}%
				\else
					{east}%
				\fi
				{\pgfkeysvalueof{/tikz/fiberkeys/size} * 0.5}
				{\pgfkeysvalueof{/tikz/fiberkeys/size}}
			\fi
			{\pgfkeysvalueof{/tikz/fiberkeys/drawbase}}%
		}
	},
	/tikz/fiberkeys/drawer/.code n args={9}{%
		\tikzset{
			fibershape,
			minimum width=#7*\NODESIZE,
			minimum height=#8*\NODESIZE,
			append after command={
				\pgfextra{\let\bdr=\tikzlastnode%
				\if#5a	
					\ifnum#9>0
						\draw[#3, #4] (\bdr.#6 west) to (\bdr.#6 east) {};
					\fi
					\node[draw=#3, #4, circle, minimum size=#2*0.5*\NODESIZE, anchor=#6] at ([xshift=-0.1*#2*\NODESIZE]\bdr.#6) () {};
					\node[draw=#3, #4, circle, minimum size=#2*0.5*\NODESIZE, anchor=#6] at (\bdr.#6) () {};
					\node[draw=#3, #4, circle, minimum size=#2*0.5*\NODESIZE, anchor=#6] at ([xshift=0.1*#2*\NODESIZE]\bdr.#6) () {};
				\fi
				\if#5b	
					\ifnum#9>0
						\draw[#3, #4] (\bdr.north #6) to (\bdr.south #6) {};
					\fi
					\node[draw=#3, #4, circle, minimum size=#2*0.5*\NODESIZE, anchor=#6] at ([yshift=0.1*#2*\NODESIZE]\bdr.#6) () {};
					\node[draw=#3, #4, circle, minimum size=#2*0.5*\NODESIZE, anchor=#6] at (\bdr.#6) () {};
					\node[draw=#3, #4, circle, minimum size=#2*0.5*\NODESIZE, anchor=#6] at ([yshift=-0.1*#2*\NODESIZE]\bdr.#6) () {};
				\fi
				}
			}
		}
	},
}
\newcount\portcount

\pgfdeclareshape{fiberswitchshape}{
	\savedmacro\nin{
		\edef\nin{\pgfkeysvalueof{/tikz/fiberswitchkeys/nin}}%
	}
	\savedmacro\nout{
		\edef\nout{\pgfkeysvalueof{/tikz/fiberswitchkeys/nout}}%
	}
	\savedmacro\direction{
		\edef\direction{\pgfkeysvalueof{/tikz/fiberswitchkeys/direction}}%
	}
	\inheritsavedanchors[from=basic] 
	
	\inheritanchor[from=basic]{center}
	\inheritanchor[from=basic]{mid}		
	\inheritanchor[from=basic]{base}	
	\inheritanchor[from=basic]{north}
	\inheritanchor[from=basic]{south}		
	\inheritanchor[from=basic]{west}		
	\inheritanchor[from=basic]{mid west}				
	\inheritanchor[from=basic]{base west}		
	\inheritanchor[from=basic]{north west}		
	\inheritanchor[from=basic]{south west}		
	\inheritanchor[from=basic]{east}
	\inheritanchor[from=basic]{mid east}
	\inheritanchor[from=basic]{base east}	
	\inheritanchor[from=basic]{north east}
	\inheritanchor[from=basic]{south east}
	\inheritanchor[from=basic]{west south west}%
	\inheritanchor[from=basic]{west north west}%
	\inheritanchor[from=basic]{east north east}%
	\inheritanchor[from=basic]{east south east}%
	\inheritanchor[from=basic]{north north west}%
	\inheritanchor[from=basic]{north north east}%
	\inheritanchor[from=basic]{south south west}%
	\inheritanchor[from=basic]{south south east}%
	
	\inheritanchorborder[from=basic]	
	\inheritbackgroundpath[from=basic]

    \pgfutil@g@addto@macro\pgf@sh@s@fiberswitchshape{%
        \pgfmathsetcount{\portcount}{0}
        \pgfmathloop%
        \ifnum\the\portcount<\nin
	        \pgfutil@ifundefined{pgf@anchor@fiberswitchshape@in\the\portcount}{
		        \expandafter\xdef\csname pgf@anchor@fiberswitchshape@in\the\portcount\endcsname{%
		            \noexpand\fiberswitchshape@port[\the\portcount]{0}
		        }%
		    }{}%
	        \ifnum\the\portcount=0
    		    \pgfutil@ifundefined{pgf@anchor@fiberswitchshape@in}{%
		        \expandafter\xdef\csname pgf@anchor@fiberswitchshape@in\endcsname{%
		            \noexpand\fiberswitchshape@port[\the\portcount]{0}
		        }%
		        }{}%
		    \fi
	        \pgfmathaddtocount{\portcount}{1}	
	        \repeatpgfmathloop
        \pgfmathsetcount{\portcount}{0}
        \pgfmathloop%
    	\ifnum\the\portcount<\nout
	        \pgfutil@ifundefined{pgf@anchor@fiberswitchshape@out\the\portcount}{%
		        \expandafter\xdef\csname pgf@anchor@fiberswitchshape@out\the\portcount\endcsname{%
		            \noexpand\fiberswitchshape@port[\the\portcount]{1}
		        }%
		    }{}%
	        \ifnum\the\portcount=0
    		    \pgfutil@ifundefined{pgf@anchor@fiberswitchshape@out}{%
		        \expandafter\xdef\csname pgf@anchor@fiberswitchshape@out\endcsname{%
		            \noexpand\fiberswitchshape@port[\the\portcount]{1}
		        }%
		        }{}%
		    \fi
	        \pgfmathaddtocount{\portcount}{1}	
	        \repeatpgfmathloop
	}
}

\def\fiberswitchshape@port[#1]#2{
    \northeast \pgf@xa=\pgf@x \pgf@ya=\pgf@y
    \southwest \pgf@xb=\pgf@x \pgf@yb=\pgf@y
    
    \ifnum#2=0	
	    \if\direction\direce	
	    	\pgf@x=\pgf@xb
		    \pgf@yc=\pgf@ya \advance\pgf@yc by -\pgf@yb	
		    \pgfmathsetlength{\pgf@y}{\pgf@ya-(#1 + 0.5)*(\pgf@yc/\nin)}%
	    \fi
	    \if\direction\direcw
	    	\pgf@x=\pgf@xa
		    \pgf@yc=\pgf@ya \advance\pgf@yc by -\pgf@yb	
		    \pgfmathsetlength{\pgf@y}{\pgf@ya-(#1 + 0.5)*(\pgf@yc/\nin)}%
	    \fi
	    \if\direction\direcn
	    	\pgf@y=\pgf@yb
		    \pgf@xc=\pgf@xa \advance\pgf@xc by -\pgf@xb	
		    \pgfmathsetlength{\pgf@x}{\pgf@xb+(#1 + 0.5)*(\pgf@xc/\nin)}%
	    \fi
	    \if\direction\direcs
	    	\pgf@y=\pgf@ya
		    \pgf@xc=\pgf@xa \advance\pgf@xc by -\pgf@xb	
		    \pgfmathsetlength{\pgf@x}{\pgf@xb+(#1 + 0.5)*(\pgf@xc/\nin)}%
	    \fi
	\else	
	    \if\direction\direce	
	    	\pgf@x=\pgf@xa
		    \pgf@yc=\pgf@ya \advance\pgf@yc by -\pgf@yb	
		    \pgfmathsetlength{\pgf@y}{\pgf@ya-(#1 + 0.5)*(\pgf@yc/\nout)}%
	    \fi
	    \if\direction\direcw
	    	\pgf@x=\pgf@xb
		    \pgf@yc=\pgf@ya \advance\pgf@yc by -\pgf@yb	
		    \pgfmathsetlength{\pgf@y}{\pgf@ya-(#1 + 0.5)*(\pgf@yc/\nout)}%
	    \fi
	    \if\direction\direcn
	    	\pgf@y=\pgf@ya
		    \pgf@xc=\pgf@xa \advance\pgf@xc by -\pgf@xb	
		    \pgfmathsetlength{\pgf@x}{\pgf@xb+(#1 + 0.5)*(\pgf@xc/\nout)}%
	    \fi
	    \if\direction\direcs
	    	\pgf@y=\pgf@yb
		    \pgf@xc=\pgf@xa \advance\pgf@xc by -\pgf@xb	
		    \pgfmathsetlength{\pgf@x}{\pgf@xb+(#1 + 0.5)*(\pgf@xc/\nout)}%
	    \fi
	\fi
}

\pgfaddtoshape{fiberswitchshape}{
	\anchorlet{c}{center}		
	\anchorlet{n}{north}
	\anchorlet{e}{east}
	\anchorlet{s}{south}
	\anchorlet{w}{west}			
	\anchorlet{se}{south east}
	\anchorlet{sw}{south west}
	\anchorlet{ne}{north east}
	\anchorlet{nw}{north west}
	\anchorlet{wsw}{west south west}
	\anchorlet{wnw}{west north west}
	\anchorlet{ene}{east north east}
	\anchorlet{ese}{east south east}
	\anchorlet{nnw}{north north west}
	\anchorlet{nne}{north north east}
	\anchorlet{ssw}{south south west}
	\anchorlet{sse}{south south east}
}

\tikzset{
	/tikz/fiberswitchkeys/.cd,
	size/.initial=1,
	color/.initial=O,
	direction/.initial=e,
	linestyle/.initial={linestyle, inner sep=0.5mm},
	nin/.initial=1,	
	nout/.initial=3,
	/tikz/fiberswitch/.code={
		\pgfqkeys{/tikz/fiberswitchkeys}{#1}%
		\tikzset{/tikz/fiberswitchkeys/drawer/.expanded=%
			{\pgfkeysvalueof{/tikz/fiberswitchkeys/size}}%
			{\pgfkeysvalueof{/tikz/fiberswitchkeys/color}}%
			{\pgfkeysvalueof{/tikz/fiberswitchkeys/linestyle}}%
			{\pgfkeysvalueof{/tikz/fiberswitchkeys/nout}}%
			\if\pgfkeysvalueof{/tikz/fiberswitchkeys/direction}e
				{0}%
			\fi
			\if\pgfkeysvalueof{/tikz/fiberswitchkeys/direction}w
				{0}%
			\fi
			\if\pgfkeysvalueof{/tikz/fiberswitchkeys/direction}n
				{1}%
			\fi
			\if\pgfkeysvalueof{/tikz/fiberswitchkeys/direction}s
				{1}%
			\fi
			{\pgfkeysvalueof{/tikz/fiberswitchkeys/direction}}%
		}
	},
	/tikz/fiberswitchkeys/drawer/.code n args={6}{%
		\tikzset{
			fiberswitchshape,
			draw,
			minimum height = #1*\NODESIZE,
			minimum width = #1*\NODESIZE,
			#2,
			#3,
			append after command={
				\pgfextra{\let\bdr=\tikzlastnode%
						\ifnum#5>0
							\node[coordinate] at ($(\bdr.in)!0.25!(\bdr.out0 -| \bdr.in)$) (circlein){};
							\foreach \n [evaluate=\n as \nport using int(\n-1)] in {1,...,#4}{
								\node[coordinate] at ($(\bdr.out\nport)!0.25!(\bdr.in -| \bdr.out\nport)$) (circleout\nport){};
							}
						\else
							\node[coordinate] at ($(\bdr.in)!0.25!(\bdr.out0 |- \bdr.in)$) (circlein){};
							\foreach \n [evaluate=\n as \nport using int(\n-1)] in {1,...,#4}{
								\node[coordinate] at ($(\bdr.out\nport)!0.25!(\bdr.in |- \bdr.out\nport)$) (circleout\nport){};
							}
						\fi

						\draw[---, #2, #3, fill] (\bdr.in) to (circlein) circle (0.05);
						\foreach \n [evaluate=\n as \nport using int(\n-1)] in {1,...,#4}{
							\draw[---, #2, #3, fill] (\bdr.out\nport) to (circleout\nport) circle (0.05);
						}

						\tikzmath{
							int \nportmax;
							\nportmax = int(#4-1);
						}
						\draw[---, #2, #3] (circlein) to (circleout0){};
						\if#6e
							\draw[-->, #2, #3, looseness=0.8] ($(circlein)!0.6!(circleout0)$) to [out=-60, in=60]($(circlein)!0.6!(circleout\nportmax)$) {};
						\fi
						\if#6w
							\draw[-->, #2, #3, looseness=0.8] ($(circlein)!0.6!(circleout0)$) to [out=-120, in=120]($(circlein)!0.6!(circleout\nportmax)$) {};
						\fi
						\if#6s
							\draw[-->, #2, #3, looseness=0.8] ($(circlein)!0.6!(circleout0)$) to [out=-30, in=-150]($(circlein)!0.6!(circleout\nportmax)$) {};
						\fi
						\if#6n
							\draw[-->, #2, #3, looseness=0.8] ($(circlein)!0.6!(circleout0)$) to [out=30, in=150]($(circlein)!0.6!(circleout\nportmax)$) {};
						\fi

				}
			}
		}
	},
}
\pgfdeclareshape{filtershape}{
	\savedmacro\direction{
		\edef\direction{\pgfkeysvalueof{/tikz/filterkeys/direction}}%
	}
	\saveddimen\minwidth{
		\pgfmathsetlength\pgf@x{\pgfshapeminwidth}%
	}
	\saveddimen\minheight{
		\pgfmathsetlength\pgf@x{\pgfshapeminheight}%
	}
	\inheritsavedanchors[from=basic]

	\inheritanchor[from=basic]{center}
	\inheritanchor[from=basic]{mid}
	\inheritanchor[from=basic]{base}
	\inheritanchor[from=basic]{north}
	\inheritanchor[from=basic]{south}
	\inheritanchor[from=basic]{west}
	\inheritanchor[from=basic]{mid west}
	\inheritanchor[from=basic]{base west}
	\inheritanchor[from=basic]{north west}
	\inheritanchor[from=basic]{south west}
	\inheritanchor[from=basic]{east}
	\inheritanchor[from=basic]{mid east}
	\inheritanchor[from=basic]{base east}
	\inheritanchor[from=basic]{north east}
	\inheritanchor[from=basic]{south east}
	\inheritanchor[from=basic]{west south west}%
	\inheritanchor[from=basic]{west north west}%
	\inheritanchor[from=basic]{east north east}%
	\inheritanchor[from=basic]{east south east}%
	\inheritanchor[from=basic]{north north west}%
	\inheritanchor[from=basic]{north north east}%
	\inheritanchor[from=basic]{south south west}%
	\inheritanchor[from=basic]{south south east}%

	\inheritanchorborder[from=basic]
	\inheritbackgroundpath[from=basic]

	\pgfutil@g@addto@macro\pgf@sh@s@filtershape{%
		\pgfutil@ifundefined{pgf@anchor@filtershape@in0}{
			\expandafter\xdef\csname pgf@anchor@filtershape@in0\endcsname{%
				\noexpand\filtershape@port{0}
			}%
		}{}%
		\pgfutil@ifundefined{pgf@anchor@filtershape@in}{
			\expandafter\xdef\csname pgf@anchor@filtershape@in\endcsname{%
				\noexpand\filtershape@port{0}
			}%
		}{}%
		\pgfutil@ifundefined{pgf@anchor@filtershape@out0}{
			\expandafter\xdef\csname pgf@anchor@filtershape@out0\endcsname{%
				\noexpand\filtershape@port{1}
			}%
		}{}%
		\pgfutil@ifundefined{pgf@anchor@filtershape@out}{
			\expandafter\xdef\csname pgf@anchor@filtershape@out\endcsname{%
				\noexpand\filtershape@port{1}
			}%
		}{}%
	}
}

\def\filtershape@port#1{
	\northeast	

	\ifnum#1=0	
		\if\direction\direce
			\pgf@x=-\pgf@x
			\pgf@ya= \pgf@y
			\pgfmathsetlength{\pgf@y}{\pgf@ya-0.5*\minheight}%
		\fi
		\if\direction\direcw
			\pgf@x=\pgf@x
			\pgf@ya= \pgf@y
			\pgfmathsetlength{\pgf@y}{\pgf@ya-0.5*\minheight}%
		\fi
		\if\direction\direcn
			\pgf@y=-\pgf@y
			\pgf@xa=\pgf@x
			\pgfmathsetlength{\pgf@x}{\pgf@xa-0.5*\minwidth}%
		\fi
		\if\direction\direcs
			\pgf@y=\pgf@y
			\pgf@xa= \pgf@x
			\pgfmathsetlength{\pgf@x}{\pgf@xa-0.5*\minwidth}%
		\fi
	\else	
		\if\direction\direce
			\pgf@x=\pgf@x
			\pgf@ya= \pgf@y
			\pgfmathsetlength{\pgf@y}{\pgf@ya-0.5*\minheight}%
		\fi
		\if\direction\direcw
			\pgf@x=-\pgf@x
			\pgf@ya= \pgf@y
			\pgfmathsetlength{\pgf@y}{\pgf@ya-0.5*\minheight}%
		\fi
		\if\direction\direcn
			\pgf@y=\pgf@y
			\pgf@xa= \pgf@x
			\pgfmathsetlength{\pgf@x}{\pgf@xa-0.5*\minwidth}%
		\fi
		\if\direction\direcs
			\pgf@y=-\pgf@y
			\pgf@xa= \pgf@x
			\pgfmathsetlength{\pgf@x}{\pgf@xa-0.5*\minwidth}%
		\fi
	\fi
}

\pgfaddtoshape{filtershape}{
	\anchorlet{c}{center}
	\anchorlet{n}{north}
	\anchorlet{e}{east}
	\anchorlet{s}{south}
	\anchorlet{w}{west}
	\anchorlet{se}{south east}
	\anchorlet{sw}{south west}
	\anchorlet{ne}{north east}
	\anchorlet{nw}{north west}
	\anchorlet{wsw}{west south west}
	\anchorlet{wnw}{west north west}
	\anchorlet{ene}{east north east}
	\anchorlet{ese}{east south east}
	\anchorlet{nnw}{north north west}
	\anchorlet{nne}{north north east}
	\anchorlet{ssw}{south south west}
	\anchorlet{sse}{south south east}
}

\pgfmathsetmacro{\WSSSINEHEIGHT}{0.06}

\tikzset{
	/tikz/filterkeys/.cd,
	size/.initial=0.5,
	color/.initial=O,
	direction/.initial=e,
	linestyle/.initial={linestyle},
	fillgradient/.initial=O,
	/tikz/filter/.code={
			\pgfqkeys{/tikz/filterkeys}{#1}%
			\tikzset{/tikz/filterkeys/drawer/.expanded=%
					{\pgfkeysvalueof{/tikz/filterkeys/direction}}%
					{\pgfkeysvalueof{/tikz/filterkeys/size}}%
					{\pgfkeysvalueof{/tikz/filterkeys/color}}%
					{\pgfkeysvalueof{/tikz/filterkeys/linestyle}}%
					{\pgfkeysvalueof{/tikz/filterkeys/fillgradient}}%
			}
		},
	/tikz/filterkeys/drawer/.code n args={5}{%
			\tikzset{
				filtershape,
				minimum height=#2*\NODESIZE,
				minimum width=#2*\NODESIZE,
				#3,
				#4,
				draw,
				append after command={
						\pgfextra{\let\bdr=\tikzlastnode%
							\node[#5, fit=(\bdr.nw)(\bdr.se)] (boxgradient){};

							\node[coordinate] at (\bdr.wnw -| \bdr.nnw) (hl){};
							\node[coordinate] at (\bdr.ene -| \bdr.nne) (hr){};
							\draw[#3,---, rounded corners = 0] (hl) sin ($(hl)!0.25!(hr) + (0,0.002*#2*\NODESIZE)$) cos ($(hl)!0.5!(hr)$) sin ($(hl)!0.75!(hr) + (0,-0.002*#2*\NODESIZE)$) cos (hr);

							\node[coordinate] at (\bdr.w -| \bdr.nnw) (ml){};
							\node[coordinate] at (\bdr.e -| \bdr.nne) (mr){};
							\draw[#3,---, rounded corners = 0] (ml) sin ($(ml)!0.25!(mr) + (0,0.002*#2*\NODESIZE)$) cos ($(ml)!0.5!(mr)$) sin ($(ml)!0.75!(mr) + (0,-0.002*#2*\NODESIZE)$) cos (mr);

							\node[coordinate] at (\bdr.wsw -| \bdr.nnw) (ll){};
							\node[coordinate] at (\bdr.ese -| \bdr.nne) (lr){};
							\draw[#3,---, rounded corners = 0] (ll) sin ($(ll)!0.25!(lr) + (0,0.002*#2*\NODESIZE)$) cos ($(ll)!0.5!(lr)$) sin ($(ll)!0.75!(lr) + (0,-0.002*#2*\NODESIZE)$) cos (lr);

							\draw[#3, ---] ($(hl)!0.25!(hr) - (0,0.002*#2*\NODESIZE)$) -- ($(hl)!0.75!(hr) + (0,0.002*#2*\NODESIZE)$);
							\draw[#3, ---] ($(ll)!0.25!(lr) - (0,0.002*#2*\NODESIZE)$) -- ($(ll)!0.75!(lr) + (0,0.002*#2*\NODESIZE)$);
						}
					}
			}
		},
}
\newcount\portcount

\pgfdeclareshape{lensshape}{
	\savedmacro\nport{
		\edef\nport{\pgfkeysvalueof{/tikz/lenskeys/nport}}%
	}
	\inheritsavedanchors[from=basic] 
	
	\inheritanchor[from=basic]{center}
	\inheritanchor[from=basic]{mid}		
	\inheritanchor[from=basic]{base}	
	\inheritanchor[from=basic]{north}
	\inheritanchor[from=basic]{south}		
	\inheritanchor[from=basic]{west}		
	\inheritanchor[from=basic]{mid west}				
	\inheritanchor[from=basic]{base west}		
	\inheritanchor[from=basic]{north west}		
	\inheritanchor[from=basic]{south west}		
	\inheritanchor[from=basic]{east}
	\inheritanchor[from=basic]{mid east}
	\inheritanchor[from=basic]{base east}	
	\inheritanchor[from=basic]{north east}
	\inheritanchor[from=basic]{south east}
	\inheritanchor[from=basic]{west south west}%
	\inheritanchor[from=basic]{west north west}%
	\inheritanchor[from=basic]{east north east}%
	\inheritanchor[from=basic]{east south east}%
	\inheritanchor[from=basic]{north north west}%
	\inheritanchor[from=basic]{north north east}%
	\inheritanchor[from=basic]{south south west}%
	\inheritanchor[from=basic]{south south east}%
	
	\inheritanchorborder[from=basic]	
	\inheritbackgroundpath[from=basic]

    \pgfutil@g@addto@macro\pgf@sh@s@lensshape{%
        \pgfmathsetcount{\portcount}{0}
        \pgfmathloop%
        \ifnum\the\portcount<\nport
	        \pgfutil@ifundefined{pgf@anchor@lensshape@in\the\portcount}{
		        \expandafter\xdef\csname pgf@anchor@lensshape@in\the\portcount\endcsname{%
		            \noexpand\lensshape@port[\the\portcount]
		        }%
		    }{}%
	        \ifnum\the\portcount=0
    		    \pgfutil@ifundefined{pgf@anchor@lensshape@in}{%
		        \expandafter\xdef\csname pgf@anchor@lensshape@in\endcsname{%
		            \noexpand\lensshape@port[\the\portcount]
		        }%
		        }{}%
		    \fi
		    \pgfutil@ifundefined{pgf@anchor@lensshape@out\the\portcount}{%
		        \expandafter\xdef\csname pgf@anchor@lensshape@out\the\portcount\endcsname{%
		            \noexpand\lensshape@port[\the\portcount]
		        }%
		    }{}%
	        \ifnum\the\portcount=0
    		    \pgfutil@ifundefined{pgf@anchor@lensshape@out}{%
		        \expandafter\xdef\csname pgf@anchor@lensshape@out\endcsname{%
		            \noexpand\lensshape@port[\the\portcount]
		        }%
		        }{}%
		    \fi
	        \pgfmathaddtocount{\portcount}{1}	
	        \repeatpgfmathloop
	}
}

\def\lensshape@port[#1]{
    \northeast \pgf@xa=\pgf@x \pgf@ya=\pgf@y
    \southwest \pgf@xb=\pgf@x \pgf@yb=\pgf@y
    
	\pgfmathsetlength{\pgf@x}{0.5*\pgf@xa + 0.5*\pgf@xb}%
    \pgf@yc=\pgf@ya \advance\pgf@yc by -\pgf@yb	
    \pgfmathsetlength{\pgf@y}{\pgf@ya-(#1 + 0.5)*(\pgf@yc/\nport)}%
}

\pgfaddtoshape{lensshape}{
	\anchorlet{c}{center}		
	\anchorlet{n}{north}
	\anchorlet{e}{east}
	\anchorlet{s}{south}
	\anchorlet{w}{west}			
	\anchorlet{se}{south east}
	\anchorlet{sw}{south west}
	\anchorlet{ne}{north east}
	\anchorlet{nw}{north west}
	\anchorlet{wsw}{west south west}
	\anchorlet{wnw}{west north west}
	\anchorlet{ene}{east north east}
	\anchorlet{ese}{east south east}
	\anchorlet{nnw}{north north west}
	\anchorlet{nne}{north north east}
	\anchorlet{ssw}{south south west}
	\anchorlet{sse}{south south east}
}

\tikzset{
	/tikz/lenskeys/.cd,
	height/.initial=1,
	width/.initial=0.3,	
	color/.initial=O,
	rotation/.initial=0,
	linestyle/.initial={linestyle, inner sep=0.5mm},
	nport/.initial=1,
	/tikz/lens/.code={
		\pgfqkeys{/tikz/lenskeys}{#1}%
		\tikzset{/tikz/lenskeys/drawer/.expanded=%
			{\pgfkeysvalueof{/tikz/lenskeys/height}}%
			{\pgfkeysvalueof{/tikz/lenskeys/width}}%
			{\pgfkeysvalueof{/tikz/lenskeys/color}}%
			{\pgfkeysvalueof{/tikz/lenskeys/linestyle}}%
			{\pgfkeysvalueof{/tikz/lenskeys/rotation}}%
		}
	},
	/tikz/lenskeys/drawer/.code n args={5}{%
		\tikzset{
			lensshape,
			rotate=#5,
			minimum height=#1*\NODESIZE,
			minimum width=#2*\NODESIZE,
			append after command={
				\pgfextra{\let\bdr=\tikzlastnode%
					\draw[---, #3, #4, rounded corners = 0] (\bdr.n) to [in=120+#5, out=240+#5] (\bdr.s) to [in=-60+#5, out=60+#5] (\bdr.n) -- cycle;
				}
			}
		}
	},
}

\newcount\portcount

\pgfdeclareshape{mirrorshape}{
	\savedmacro\nport{
		\edef\nport{\pgfkeysvalueof{/tikz/mirrorkeys/nport}}%
	}
	\inheritsavedanchors[from=basic] 
	
	\inheritanchor[from=basic]{center}
	\inheritanchor[from=basic]{mid}		
	\inheritanchor[from=basic]{base}	
	\inheritanchor[from=basic]{north}
	\inheritanchor[from=basic]{south}		
	\inheritanchor[from=basic]{west}		
	\inheritanchor[from=basic]{mid west}				
	\inheritanchor[from=basic]{base west}		
	\inheritanchor[from=basic]{north west}		
	\inheritanchor[from=basic]{south west}		
	\inheritanchor[from=basic]{east}
	\inheritanchor[from=basic]{mid east}
	\inheritanchor[from=basic]{base east}	
	\inheritanchor[from=basic]{north east}
	\inheritanchor[from=basic]{south east}
	\inheritanchor[from=basic]{west south west}%
	\inheritanchor[from=basic]{west north west}%
	\inheritanchor[from=basic]{east north east}%
	\inheritanchor[from=basic]{east south east}%
	\inheritanchor[from=basic]{north north west}%
	\inheritanchor[from=basic]{north north east}%
	\inheritanchor[from=basic]{south south west}%
	\inheritanchor[from=basic]{south south east}%
	
	\inheritanchorborder[from=basic]	
	\inheritbackgroundpath[from=basic]

    \pgfutil@g@addto@macro\pgf@sh@s@mirrorshape{%
        \pgfmathsetcount{\portcount}{0}
        \pgfmathloop%
        \ifnum\the\portcount<\nport
	        \pgfutil@ifundefined{pgf@anchor@mirrorshape@in\the\portcount}{
		        \expandafter\xdef\csname pgf@anchor@mirrorshape@in\the\portcount\endcsname{%
		            \noexpand\mirrorshape@port[\the\portcount]
		        }%
		    }{}%
	        \ifnum\the\portcount=0
    		    \pgfutil@ifundefined{pgf@anchor@mirrorshape@in}{%
		        \expandafter\xdef\csname pgf@anchor@mirrorshape@in\endcsname{%
		            \noexpand\mirrorshape@port[\the\portcount]
		        }%
		        }{}%
		    \fi
		    \pgfutil@ifundefined{pgf@anchor@mirrorshape@out\the\portcount}{%
		        \expandafter\xdef\csname pgf@anchor@mirrorshape@out\the\portcount\endcsname{%
		            \noexpand\mirrorshape@port[\the\portcount]
		        }%
		    }{}%
	        \ifnum\the\portcount=0
    		    \pgfutil@ifundefined{pgf@anchor@mirrorshape@out}{%
		        \expandafter\xdef\csname pgf@anchor@mirrorshape@out\endcsname{%
		            \noexpand\mirrorshape@port[\the\portcount]
		        }%
		        }{}%
		    \fi
	        \pgfmathaddtocount{\portcount}{1}	
	        \repeatpgfmathloop
	}
}

\def\mirrorshape@port[#1]{
    \northeast \pgf@xa=\pgf@x \pgf@ya=\pgf@y
    \southwest \pgf@xb=\pgf@x \pgf@yb=\pgf@y
    
	\pgf@x=\pgf@xb
    \pgf@yc=\pgf@ya \advance\pgf@yc by -\pgf@yb	
    \pgfmathsetlength{\pgf@y}{\pgf@ya-(#1 + 0.5)*(\pgf@yc/\nport)}%
}

\pgfaddtoshape{mirrorshape}{
	\anchorlet{c}{center}		
	\anchorlet{n}{north}
	\anchorlet{e}{east}
	\anchorlet{s}{south}
	\anchorlet{w}{west}			
	\anchorlet{se}{south east}
	\anchorlet{sw}{south west}
	\anchorlet{ne}{north east}
	\anchorlet{nw}{north west}
	\anchorlet{wsw}{west south west}
	\anchorlet{wnw}{west north west}
	\anchorlet{ene}{east north east}
	\anchorlet{ese}{east south east}
	\anchorlet{nnw}{north north west}
	\anchorlet{nne}{north north east}
	\anchorlet{ssw}{south south west}
	\anchorlet{sse}{south south east}
}

\tikzset{
	/tikz/mirrorkeys/.cd,
	height/.initial=1,
	width/.initial=0.15,	
	color/.initial=O,
	rotation/.initial=0,
	linestyle/.initial={linestyle, inner sep=0.5mm},
	nport/.initial=1,
	nlines/.initial=5,
	/tikz/mirror/.code={
		\pgfqkeys{/tikz/mirrorkeys}{#1}%
		\tikzset{/tikz/mirrorkeys/drawer/.expanded=%
			{\pgfkeysvalueof{/tikz/mirrorkeys/height}}%
			{\pgfkeysvalueof{/tikz/mirrorkeys/width}}%
			{\pgfkeysvalueof{/tikz/mirrorkeys/color}}%
			{\pgfkeysvalueof{/tikz/mirrorkeys/linestyle}}%
			{\pgfkeysvalueof{/tikz/mirrorkeys/rotation}}%
			{\pgfkeysvalueof{/tikz/mirrorkeys/nlines}}%
		}
	},
	/tikz/mirrorkeys/drawer/.code n args={6}{%
		\tikzset{
			mirrorshape,
			rotate=#5,
			minimum height=#1*\NODESIZE,
			minimum width=#2*\NODESIZE,
			append after command={
				\pgfextra{\let\bdr=\tikzlastnode%
					\draw[---, #3, #4] (\bdr.nw) to (\bdr.sw){};
					\foreach \nline [evaluate=\nline as \linepos using (\nline-0.8)/(#6-0.6)] in {1,...,#6}{
						\draw[---, #3, #4] ($(\bdr.nw)!\linepos!(\bdr.sw)$) to +(#5-45:#2*1.41421356237*\NODESIZE pt){};
					}
				}
			}
		}
	},
}
\pgfdeclareshape{muxshape}{
	\savedmacro\nports{
		\edef\nports{\pgfkeysvalueof{/tikz/muxkeys/nports}}%
	}
	\savedmacro\direction{
		\edef\direction{\pgfkeysvalueof{/tikz/muxkeys/direction}}%
	}
	\savedmacro\inverted{
		\edef\inverted{\pgfkeysvalueof{/tikz/muxkeys/inverted}}%
	}
	\savedmacro\ninports{
		\ifnum\inverted=0
			\edef\ninports{\nports}
		\else
			\edef\ninports{1}%
		\fi
	}
	\savedmacro\noutports{
		\ifnum\inverted=0
			\edef\noutports{1}%
		\else
			\edef\noutports{\nports}%
		\fi
	}
	\inheritsavedanchors[from=basic]

	\inheritanchor[from=basic]{center}
	\inheritanchor[from=basic]{mid}
	\inheritanchor[from=basic]{base}
	\inheritanchor[from=basic]{north}
	\inheritanchor[from=basic]{south}
	\inheritanchor[from=basic]{west}
	\inheritanchor[from=basic]{mid west}
	\inheritanchor[from=basic]{base west}
	\inheritanchor[from=basic]{north west}
	\inheritanchor[from=basic]{south west}
	\inheritanchor[from=basic]{east}
	\inheritanchor[from=basic]{mid east}
	\inheritanchor[from=basic]{base east}
	\inheritanchor[from=basic]{north east}
	\inheritanchor[from=basic]{south east}
	\inheritanchor[from=basic]{west south west}%
	\inheritanchor[from=basic]{west north west}%
	\inheritanchor[from=basic]{east north east}%
	\inheritanchor[from=basic]{east south east}%
	\inheritanchor[from=basic]{north north west}%
	\inheritanchor[from=basic]{north north east}%
	\inheritanchor[from=basic]{south south west}%
	\inheritanchor[from=basic]{south south east}%

	\inheritanchorborder[from=basic]
	\inheritbackgroundpath[from=basic]

	\pgfutil@g@addto@macro\pgf@sh@s@muxshape{%
		\pgfmathsetcount{\portcount}{0}
		\pgfmathloop%
		\ifnum\the\portcount<\nports
		\ifnum\the\portcount<\ninports
			\pgfutil@ifundefined{pgf@anchor@muxshape@in\the\portcount}{
				\expandafter\xdef\csname pgf@anchor@muxshape@in\the\portcount\endcsname{%
					\noexpand\muxshape@port[\the\portcount]{0}
				}%
			}{}%
			\ifnum\the\portcount=0
				\pgfutil@ifundefined{pgf@anchor@muxshape@in}{%
					\expandafter\xdef\csname pgf@anchor@muxshape@in\endcsname{%
						\noexpand\muxshape@port[\the\portcount]{0}
					}%
				}{}%
			\fi
		\fi
		\ifnum\the\portcount<\noutports
			\pgfutil@ifundefined{pgf@anchor@muxshape@out\the\portcount}{%
				\expandafter\xdef\csname pgf@anchor@muxshape@out\the\portcount\endcsname{%
					\noexpand\muxshape@port[\the\portcount]{1}
				}%
			}{}%
			\ifnum\the\portcount=0
				\pgfutil@ifundefined{pgf@anchor@muxshape@out}{%
					\expandafter\xdef\csname pgf@anchor@muxshape@out\endcsname{%
						\noexpand\muxshape@port[\the\portcount]{1}
					}%
				}{}%
			\fi
		\fi
		\pgfmathaddtocount{\portcount}{1}	
		\repeatpgfmathloop
	}
}

\def\muxshape@port[#1]#2{
	\northeast \pgf@xa=\pgf@x \pgf@ya=\pgf@y
	\southwest \pgf@xb=\pgf@x \pgf@yb=\pgf@y

	\ifnum#2=0
		\ifnum\inverted=0
			\def\chooseports{0}
		\else
			\def\chooseports{1}
		\fi
	\else
		\ifnum\inverted=0
			\def\chooseports{1}
		\else
			\def\chooseports{0}
		\fi
	\fi

	\ifnum\chooseports=0	
		\if\direction\direce
			\pgf@x=\pgf@xb
			\pgf@yc=\pgf@ya \advance\pgf@yc by -\pgf@yb	
			\pgfmathsetlength{\pgf@y}{\pgf@ya-(#1 + 0.5)*(\pgf@yc/\nports)}%
		\fi
		\if\direction\direcw
			\pgf@x=\pgf@xa
			\pgf@yc=\pgf@ya \advance\pgf@yc by -\pgf@yb	
			\pgfmathsetlength{\pgf@y}{\pgf@ya-(#1 + 0.5)*(\pgf@yc/\nports)}%
		\fi
		\if\direction\direcn
			\pgf@y=\pgf@yb
			\pgf@xc=\pgf@xa \advance\pgf@xc by -\pgf@xb	
			\pgfmathsetlength{\pgf@x}{\pgf@xb+(#1 + 0.5)*(\pgf@xc/\nports)}%
		\fi
		\if\direction\direcs
			\pgf@y=\pgf@ya
			\pgf@xc=\pgf@xa \advance\pgf@xc by -\pgf@xb	
			\pgfmathsetlength{\pgf@x}{\pgf@xb+(#1 + 0.5)*(\pgf@xc/\nports)}%
		\fi
	\else	
		\if\direction\direce
			\pgf@x=\pgf@xa
			\pgf@yc=\pgf@ya \advance\pgf@yc by -\pgf@yb	
			\pgfmathsetlength{\pgf@y}{\pgf@ya-0.5\pgf@yc}%
		\fi
		\if\direction\direcw
			\pgf@x=\pgf@xb
			\pgf@yc=\pgf@ya \advance\pgf@yc by -\pgf@yb	
			\pgfmathsetlength{\pgf@y}{\pgf@ya-0.5\pgf@yc}%
		\fi
		\if\direction\direcn
			\pgf@y=\pgf@ya
			\pgf@xc=\pgf@xa \advance\pgf@xc by -\pgf@xb	
			\pgfmathsetlength{\pgf@x}{\pgf@xa-0.5\pgf@xc}%
		\fi
		\if\direction\direcs
			\pgf@y=\pgf@yb
			\pgf@xc=\pgf@xa \advance\pgf@xc by -\pgf@xb	
			\pgfmathsetlength{\pgf@x}{\pgf@xa-0.5\pgf@xc}%
		\fi
	\fi
}

\pgfaddtoshape{muxshape}{
	\anchorlet{c}{center}
	\anchorlet{n}{north}
	\anchorlet{e}{east}
	\anchorlet{s}{south}
	\anchorlet{w}{west}
	\anchorlet{se}{south east}
	\anchorlet{sw}{south west}
	\anchorlet{ne}{north east}
	\anchorlet{nw}{north west}
	\anchorlet{wsw}{west south west}
	\anchorlet{wnw}{west north west}
	\anchorlet{ene}{east north east}
	\anchorlet{ese}{east south east}
	\anchorlet{nnw}{north north west}
	\anchorlet{nne}{north north east}
	\anchorlet{ssw}{south south west}
	\anchorlet{sse}{south south east}
}

\tikzset{
/tikz/muxkeys/.cd,
height/.initial=1,
width/.initial=0.5,
color/.initial=O,
direction/.initial=e,
linestyle/.initial={linestyle, rounded corners = 0},
nports/.initial=3,
inverted/.initial=0,	
smux/.initial=0,
angle/.initial=60,
fillgradient/.initial=O,
/tikz/mux/.code={
\pgfqkeys{/tikz/muxkeys}{#1}%
\tikzset{/tikz/muxkeys/drawer/.expanded=%
\if\pgfkeysvalueof{/tikz/muxkeys/direction}e
	{\pgfkeysvalueof{/tikz/muxkeys/width}}%
	{\pgfkeysvalueof{/tikz/muxkeys/height}}%
	{0}%
	{-90}%
\fi
\if\pgfkeysvalueof{/tikz/muxkeys/direction}w
	{\pgfkeysvalueof{/tikz/muxkeys/width}}%
	{\pgfkeysvalueof{/tikz/muxkeys/height}}%
	{0}%
	{90}%
\fi
\if\pgfkeysvalueof{/tikz/muxkeys/direction}n
	{\pgfkeysvalueof{/tikz/muxkeys/width}}%
	{\pgfkeysvalueof{/tikz/muxkeys/height}}%
	{1}%
	{0}%
\fi
\if\pgfkeysvalueof{/tikz/muxkeys/direction}s
	{\pgfkeysvalueof{/tikz/muxkeys/width}}%
	{\pgfkeysvalueof{/tikz/muxkeys/height}}%
	{1}%
	{180}%
\fi
{\pgfkeysvalueof{/tikz/muxkeys/color}}%
{\pgfkeysvalueof{/tikz/muxkeys/linestyle}}%
{\pgfkeysvalueof{/tikz/muxkeys/smux}}%
{\pgfkeysvalueof{/tikz/muxkeys/angle}}%
{\pgfkeysvalueof{/tikz/muxkeys/fillgradient}}%
}
},
/tikz/muxkeys/drawer/.code n args={9}{%
		\tikzset{
			muxshape,
			#6,
			#5,
			minimum height=
			\ifnum#3>0	
				#1*\NODESIZE
			\else
				#2*\NODESIZE
			\fi
			,minimum width=
			\ifnum#3>0
				#2*\NODESIZE
			\else
				#1*\NODESIZE
			\fi
			,append after command={
					\pgfextra{\let\bdr=\tikzlastnode%
						\node[trapezium, line width = \NODETHICKNESS, minimum height=#1*\NODESIZE, minimum width=#2*\NODESIZE, trapezium stretches=true, rotate=#4, trapezium angle=#8, inner sep=0.001mm] at (\bdr) (trap) {};	
						\ifnum#7=0
							\draw[#9, #5, #6] (trap.bottom left corner) to (trap.top left corner) to (trap.top right corner) to (trap.bottom right corner) to cycle;
						\else
							\draw[#9, #5, #6] (trap.bottom left corner) to[in=#4-90, out=#4+90] (trap.top left corner) to (trap.top right corner) to[in=#4+90,out=#4-90] (trap.bottom right corner) to cycle;
						\fi
					}
				}
		}
	},
}
\newcount\portcount

\pgfdeclareshape{polswitchshape}{
	\savedmacro\nin{
		\edef\nin{\pgfkeysvalueof{/tikz/polswitchkeys/nin}}%
	}
	\savedmacro\nout{
		\edef\nout{\pgfkeysvalueof{/tikz/polswitchkeys/nout}}%
	}
	\savedmacro\direction{
		\edef\direction{\pgfkeysvalueof{/tikz/polswitchkeys/direction}}%
	}
	\inheritsavedanchors[from=basic] 
	
	\inheritanchor[from=basic]{center}
	\inheritanchor[from=basic]{mid}		
	\inheritanchor[from=basic]{base}	
	\inheritanchor[from=basic]{north}
	\inheritanchor[from=basic]{south}		
	\inheritanchor[from=basic]{west}		
	\inheritanchor[from=basic]{mid west}				
	\inheritanchor[from=basic]{base west}		
	\inheritanchor[from=basic]{north west}		
	\inheritanchor[from=basic]{south west}		
	\inheritanchor[from=basic]{east}
	\inheritanchor[from=basic]{mid east}
	\inheritanchor[from=basic]{base east}	
	\inheritanchor[from=basic]{north east}
	\inheritanchor[from=basic]{south east}
	\inheritanchor[from=basic]{west south west}%
	\inheritanchor[from=basic]{west north west}%
	\inheritanchor[from=basic]{east north east}%
	\inheritanchor[from=basic]{east south east}%
	\inheritanchor[from=basic]{north north west}%
	\inheritanchor[from=basic]{north north east}%
	\inheritanchor[from=basic]{south south west}%
	\inheritanchor[from=basic]{south south east}%
	
	\inheritanchorborder[from=basic]	
	\inheritbackgroundpath[from=basic]

    \pgfutil@g@addto@macro\pgf@sh@s@polswitchshape{%
        \pgfmathsetcount{\portcount}{0}
        \pgfmathloop%
        \ifnum\the\portcount<\nin
	        \pgfutil@ifundefined{pgf@anchor@polswitchshape@in\the\portcount}{
		        \expandafter\xdef\csname pgf@anchor@polswitchshape@in\the\portcount\endcsname{%
		            \noexpand\polswitchshape@port[\the\portcount]{0}
		        }%
		    }{}%
	        \ifnum\the\portcount=0
    		    \pgfutil@ifundefined{pgf@anchor@polswitchshape@in}{%
		        \expandafter\xdef\csname pgf@anchor@polswitchshape@in\endcsname{%
		            \noexpand\polswitchshape@port[\the\portcount]{0}
		        }%
		        }{}%
		    \fi
	        \pgfmathaddtocount{\portcount}{1}	
	        \repeatpgfmathloop
        \pgfmathsetcount{\portcount}{0}
        \pgfmathloop%
    	\ifnum\the\portcount<\nout
	        \pgfutil@ifundefined{pgf@anchor@polswitchshape@out\the\portcount}{%
		        \expandafter\xdef\csname pgf@anchor@polswitchshape@out\the\portcount\endcsname{%
		            \noexpand\polswitchshape@port[\the\portcount]{1}
		        }%
		    }{}%
	        \ifnum\the\portcount=0
    		    \pgfutil@ifundefined{pgf@anchor@polswitchshape@out}{%
		        \expandafter\xdef\csname pgf@anchor@polswitchshape@out\endcsname{%
		            \noexpand\polswitchshape@port[\the\portcount]{1}
		        }%
		        }{}%
		    \fi
	        \pgfmathaddtocount{\portcount}{1}	
	        \repeatpgfmathloop
	}
}

\def\polswitchshape@port[#1]#2{
    \northeast \pgf@xa=\pgf@x \pgf@ya=\pgf@y
    \southwest \pgf@xb=\pgf@x \pgf@yb=\pgf@y
    
    \ifnum#2=0	
	    \if\direction\direce	
	    	\pgf@x=\pgf@xb
		    \pgf@yc=\pgf@ya \advance\pgf@yc by -\pgf@yb	
		    \pgfmathsetlength{\pgf@y}{\pgf@ya-(#1 + 0.5)*(\pgf@yc/\nin)}%
	    \fi
	    \if\direction\direcw
	    	\pgf@x=\pgf@xa
		    \pgf@yc=\pgf@ya \advance\pgf@yc by -\pgf@yb	
		    \pgfmathsetlength{\pgf@y}{\pgf@ya-(#1 + 0.5)*(\pgf@yc/\nin)}%
	    \fi
	    \if\direction\direcn
	    	\pgf@y=\pgf@yb
		    \pgf@xc=\pgf@xa \advance\pgf@xc by -\pgf@xb	
		    \pgfmathsetlength{\pgf@x}{\pgf@xb+(#1 + 0.5)*(\pgf@xc/\nin)}%
	    \fi
	    \if\direction\direcs
	    	\pgf@y=\pgf@ya
		    \pgf@xc=\pgf@xa \advance\pgf@xc by -\pgf@xb	
		    \pgfmathsetlength{\pgf@x}{\pgf@xb+(#1 + 0.5)*(\pgf@xc/\nin)}%
	    \fi
	\else	
	    \if\direction\direce	
	    	\pgf@x=\pgf@xa
		    \pgf@yc=\pgf@ya \advance\pgf@yc by -\pgf@yb	
		    \pgfmathsetlength{\pgf@y}{\pgf@ya-(#1 + 0.5)*(\pgf@yc/\nout)}%
	    \fi
	    \if\direction\direcw
	    	\pgf@x=\pgf@xb
		    \pgf@yc=\pgf@ya \advance\pgf@yc by -\pgf@yb	
		    \pgfmathsetlength{\pgf@y}{\pgf@ya-(#1 + 0.5)*(\pgf@yc/\nout)}%
	    \fi
	    \if\direction\direcn
	    	\pgf@y=\pgf@ya
		    \pgf@xc=\pgf@xa \advance\pgf@xc by -\pgf@xb	
		    \pgfmathsetlength{\pgf@x}{\pgf@xb+(#1 + 0.5)*(\pgf@xc/\nout)}%
	    \fi
	    \if\direction\direcs
	    	\pgf@y=\pgf@yb
		    \pgf@xc=\pgf@xa \advance\pgf@xc by -\pgf@xb	
		    \pgfmathsetlength{\pgf@x}{\pgf@xb+(#1 + 0.5)*(\pgf@xc/\nout)}%
	    \fi
	\fi
}

\pgfaddtoshape{polswitchshape}{
	\anchorlet{c}{center}		
	\anchorlet{n}{north}
	\anchorlet{e}{east}
	\anchorlet{s}{south}
	\anchorlet{w}{west}			
	\anchorlet{se}{south east}
	\anchorlet{sw}{south west}
	\anchorlet{ne}{north east}
	\anchorlet{nw}{north west}
	\anchorlet{wsw}{west south west}
	\anchorlet{wnw}{west north west}
	\anchorlet{ene}{east north east}
	\anchorlet{ese}{east south east}
	\anchorlet{nnw}{north north west}
	\anchorlet{nne}{north north east}
	\anchorlet{ssw}{south south west}
	\anchorlet{sse}{south south east}
}

\tikzset{
	/tikz/polswitchkeys/.cd,
	size/.initial=1,
	color/.initial=O,
	direction/.initial=e,
	linestyle/.initial={linestyle, inner sep=0.5mm},
	nin/.initial=1,	
	nout/.initial=1, 
	/tikz/polswitch/.code={
		\pgfqkeys{/tikz/polswitchkeys}{#1}%
		\tikzset{/tikz/polswitchkeys/drawer/.expanded=%
			{\pgfkeysvalueof{/tikz/polswitchkeys/size}}%
			{\pgfkeysvalueof{/tikz/polswitchkeys/color}}%
			{\pgfkeysvalueof{/tikz/polswitchkeys/linestyle}}%
			{\pgfkeysvalueof{/tikz/polswitchkeys/nout}}%
			\if\pgfkeysvalueof{/tikz/polswitchkeys/direction}e
				{0}%
			\fi
			\if\pgfkeysvalueof{/tikz/polswitchkeys/direction}w
				{0}%
			\fi
			\if\pgfkeysvalueof{/tikz/polswitchkeys/direction}n
				{1}%
			\fi
			\if\pgfkeysvalueof{/tikz/polswitchkeys/direction}s
				{1}%
			\fi
			{\pgfkeysvalueof{/tikz/polswitchkeys/direction}}%
		}
	},
	/tikz/polswitchkeys/drawer/.code n args={6}{%
		\tikzset{
			polswitchshape,
			draw,
			minimum height = #1*\NODESIZE,
			minimum width = #1*\NODESIZE,
			#2,
			#3,
			append after command={
				\pgfextra{\let\bdr=\tikzlastnode%
						\node[coordinate] at ($(\bdr.in)!0.25!(\bdr.out)$) (circlein){};
						\node[coordinate] at ($(\bdr.in)!0.75!(\bdr.out)$) (circleout){};

						\draw[---, #2, #3, fill] (\bdr.in) to (circlein) circle (0.05);
						\draw[---, #2, #3, fill] (\bdr.out) to (circleout) circle (0.05);

						\node[coordinate] at ($(\bdr.in)!0.6!(\bdr.out)$) (circlemiddle){};
						\ifnum#5>0
							\node[coordinate] at ($(circlemiddle)!0.5!(circlemiddle -| \bdr.e)$) (circletopcor){};
							\node[coordinate] at ($(circlemiddle)!0.5!(circlemiddle -| \bdr.w)$) (circlebotcor){};
						\else
							\node[coordinate] at ($(circlemiddle)!0.5!(circlemiddle |- \bdr.n)$) (circletopcor){};
							\node[coordinate] at ($(circlemiddle)!0.5!(circlemiddle |- \bdr.s)$) (circlebotcor){};
						\fi

						\node[draw, circle, #2, #3, minimum size=0.3*\FNODESIZE] at (circletopcor) (circletop){};
						\node[draw, circle, #2, #3, minimum size=0.3*\FNODESIZE] at (circlebotcor) (circlebot){};
						\draw[-->, #2, #3] (circletop.south) to (circletop.north){};
						\draw[-->, #2, #3] (circlebot.west) to (circlebot.east){};

						\if#6e
							\draw[-->, #2, #3] ([xshift=-0.05*\FNODESIZE]circletop.south west) to [out=-120, in=120] ([xshift=-0.05*\FNODESIZE]circlebot.north west){};
						\fi
						\if#6w
							\draw[-->, #2, #3] ([xshift=0.05*\FNODESIZE]circletop.south east) to [out=-60, in=60] ([xshift=0.05*\FNODESIZE]circlebot.north east){};
						\fi
						\if#6s
							\draw[-->, #2, #3] ([yshift=0.05*\FNODESIZE]circletop.north west) to [out=150, in=30] ([yshift=0.05*\FNODESIZE]circlebot.north east){};
						\fi
						\if#6n
							\draw[-->, #2, #3] ([yshift=-0.05*\FNODESIZE]circletop.south west) to [out=-150, in=-30] ([yshift=-0.05*\FNODESIZE]circlebot.south east){};
						\fi

				}
			}
		}
	},
}
\pgfdeclareshape{pdshape}{
	\savedmacro\direction{
		\edef\direction{\pgfkeysvalueof{/tikz/pdkeys/direction}}%
	}
	\saveddimen\minwidth{
		\pgfmathsetlength\pgf@x{\pgfshapeminwidth}%
	}
	\saveddimen\minheight{
		\pgfmathsetlength\pgf@x{\pgfshapeminheight}%
	}
	\inheritsavedanchors[from=basic]

	\inheritanchor[from=basic]{center}
	\inheritanchor[from=basic]{mid}
	\inheritanchor[from=basic]{base}
	\inheritanchor[from=basic]{north}
	\inheritanchor[from=basic]{south}
	\inheritanchor[from=basic]{west}
	\inheritanchor[from=basic]{mid west}
	\inheritanchor[from=basic]{base west}
	\inheritanchor[from=basic]{north west}
	\inheritanchor[from=basic]{south west}
	\inheritanchor[from=basic]{east}
	\inheritanchor[from=basic]{mid east}
	\inheritanchor[from=basic]{base east}
	\inheritanchor[from=basic]{north east}
	\inheritanchor[from=basic]{south east}
	\inheritanchor[from=basic]{west south west}%
	\inheritanchor[from=basic]{west north west}%
	\inheritanchor[from=basic]{east north east}%
	\inheritanchor[from=basic]{east south east}%
	\inheritanchor[from=basic]{north north west}%
	\inheritanchor[from=basic]{north north east}%
	\inheritanchor[from=basic]{south south west}%
	\inheritanchor[from=basic]{south south east}%

	\inheritanchorborder[from=basic]
	\inheritbackgroundpath[from=basic]

	\pgfutil@g@addto@macro\pgf@sh@s@pdshape{%
		\pgfutil@ifundefined{pgf@anchor@pdshape@in0}{
			\expandafter\xdef\csname pgf@anchor@pdshape@in0\endcsname{%
				\noexpand\pdshape@port{0}
			}%
		}{}%
		\pgfutil@ifundefined{pgf@anchor@pdshape@in}{
			\expandafter\xdef\csname pgf@anchor@pdshape@in\endcsname{%
				\noexpand\pdshape@port{0}
			}%
		}{}%
		\pgfutil@ifundefined{pgf@anchor@pdshape@out0}{
			\expandafter\xdef\csname pgf@anchor@pdshape@out0\endcsname{%
				\noexpand\pdshape@port{1}
			}%
		}{}%
		\pgfutil@ifundefined{pgf@anchor@pdshape@out}{
			\expandafter\xdef\csname pgf@anchor@pdshape@out\endcsname{%
				\noexpand\pdshape@port{1}
			}%
		}{}%
	}
}

\def\pdshape@port#1{
	\northeast	

	\ifnum#1=0	
		\if\direction\direce
			\pgf@x=-\pgf@x
			\pgf@ya= \pgf@y
			\pgfmathsetlength{\pgf@y}{\pgf@ya-0.5*\minheight}%
		\fi
		\if\direction\direcw
			\pgf@x=\pgf@x
			\pgf@ya= \pgf@y
			\pgfmathsetlength{\pgf@y}{\pgf@ya-0.5*\minheight}%
		\fi
		\if\direction\direcn
			\pgf@y=-\pgf@y
			\pgf@xa=\pgf@x
			\pgfmathsetlength{\pgf@x}{\pgf@xa-0.5*\minwidth}%
		\fi
		\if\direction\direcs
			\pgf@y=\pgf@y
			\pgf@xa= \pgf@x
			\pgfmathsetlength{\pgf@x}{\pgf@xa-0.5*\minwidth}%
		\fi
	\else	
		\if\direction\direce
			\pgf@x=\pgf@x
			\pgf@ya= \pgf@y
			\pgfmathsetlength{\pgf@y}{\pgf@ya-0.5*\minheight}%
		\fi
		\if\direction\direcw
			\pgf@x=-\pgf@x
			\pgf@ya= \pgf@y
			\pgfmathsetlength{\pgf@y}{\pgf@ya-0.5*\minheight}%
		\fi
		\if\direction\direcn
			\pgf@y=\pgf@y
			\pgf@xa= \pgf@x
			\pgfmathsetlength{\pgf@x}{\pgf@xa-0.5*\minwidth}%
		\fi
		\if\direction\direcs
			\pgf@y=-\pgf@y
			\pgf@xa= \pgf@x
			\pgfmathsetlength{\pgf@x}{\pgf@xa-0.5*\minwidth}%
		\fi
	\fi
}

\pgfaddtoshape{pdshape}{
	\anchorlet{c}{center}
	\anchorlet{n}{north}
	\anchorlet{e}{east}
	\anchorlet{s}{south}
	\anchorlet{w}{west}
	\anchorlet{se}{south east}
	\anchorlet{sw}{south west}
	\anchorlet{ne}{north east}
	\anchorlet{nw}{north west}
	\anchorlet{wsw}{west south west}
	\anchorlet{wnw}{west north west}
	\anchorlet{ene}{east north east}
	\anchorlet{ese}{east south east}
	\anchorlet{nnw}{north north west}
	\anchorlet{nne}{north north east}
	\anchorlet{ssw}{south south west}
	\anchorlet{sse}{south south east}
}

\tikzset{
/tikz/pdkeys/.cd,
size/.initial=0.5,
color/.initial=EO,
direction/.initial=e,
linestyle/.initial={linestyle},
fillgradient/.initial=O,
/tikz/pd/.code={
		\pgfqkeys{/tikz/pdkeys}{#1}%
		\tikzset{/tikz/pdkeys/drawer/.expanded=%
				{\pgfkeysvalueof{/tikz/pdkeys/direction}}%
				{\pgfkeysvalueof{/tikz/pdkeys/size}}%
				{\pgfkeysvalueof{/tikz/pdkeys/color}}%
				{\pgfkeysvalueof{/tikz/pdkeys/linestyle}}%
				{\pgfkeysvalueof{/tikz/pdkeys/fillgradient}}%
		}
	},
/tikz/pdkeys/drawer/.code n args={5}{%
\tikzset{
pdshape,
minimum height=#2*\NODESIZE,
minimum width=#2*\NODESIZE,
#3,
#4,
draw,
append after command={
\pgfextra{\let\bdr=\tikzlastnode%
\node[#5, fit=(\bdr.nw)(\bdr.se)] (boxgradient){};
\draw[---,#3] ($(\bdr.s)!.1!(\bdr.n)$) to ($(\bdr.s)!.9!(\bdr.n)$);
\fill[#3] ({$(\bdr.s)!.3!(\bdr.n)$} -| {$(\bdr.w)!.3!(\bdr.e)$}) to ($(\bdr.s)!.7!(\bdr.n)$) to ({$(\bdr.s)!.3!(\bdr.n)$} -| {$(\bdr.w)!.7!(\bdr.e)$}) to cycle;
\draw[---,#3] ({$(\bdr.s)!.7!(\bdr.n)$} -| {$(\bdr.w)!.35!(\bdr.e)$}) to ({$(\bdr.s)!.7!(\bdr.n)$} -| {$(\bdr.w)!.65!(\bdr.e)$});
}
}
}
},
}
\pgfdeclareshape{pbsshape}{
	\savedmacro\direction{
		\edef\direction{\pgfkeysvalueof{/tikz/pbskeys/direction}}%
	}
	\saveddimen\minwidth{
		\pgfmathsetlength\pgf@x{\pgfshapeminwidth}%
	}
	\saveddimen\minheight{
		\pgfmathsetlength\pgf@x{\pgfshapeminheight}%
	}
	\savedmacro\nport{
		\edef\nport{\pgfkeysvalueof{/tikz/pbskeys/nport}}%
	}
	\inheritsavedanchors[from=basic]

	\inheritanchor[from=basic]{center}
	\inheritanchor[from=basic]{mid}
	\inheritanchor[from=basic]{base}
	\inheritanchor[from=basic]{north}
	\inheritanchor[from=basic]{south}
	\inheritanchor[from=basic]{west}
	\inheritanchor[from=basic]{mid west}
	\inheritanchor[from=basic]{base west}
	\inheritanchor[from=basic]{north west}
	\inheritanchor[from=basic]{south west}
	\inheritanchor[from=basic]{east}
	\inheritanchor[from=basic]{mid east}
	\inheritanchor[from=basic]{base east}
	\inheritanchor[from=basic]{north east}
	\inheritanchor[from=basic]{south east}
	\inheritanchor[from=basic]{west south west}%
	\inheritanchor[from=basic]{west north west}%
	\inheritanchor[from=basic]{east north east}%
	\inheritanchor[from=basic]{east south east}%
	\inheritanchor[from=basic]{north north west}%
	\inheritanchor[from=basic]{north north east}%
	\inheritanchor[from=basic]{south south west}%
	\inheritanchor[from=basic]{south south east}%

	\inheritanchorborder[from=basic]
	\inheritbackgroundpath[from=basic]

	\pgfutil@g@addto@macro\pgf@sh@s@pbsshape{%
		\pgfutil@ifundefined{pgf@anchor@pbsshape@in0}{
			\expandafter\xdef\csname pgf@anchor@pbsshape@in0\endcsname{%
				\noexpand\pbsshape@port[0]{0}
			}%
		}{}%
		\pgfutil@ifundefined{pgf@anchor@pbsshape@in1}{
			\expandafter\xdef\csname pgf@anchor@pbsshape@in1\endcsname{%
				\noexpand\pbsshape@port[1]{0}
			}%
		}{}%
		\pgfutil@ifundefined{pgf@anchor@pbsshape@in}{
			\expandafter\xdef\csname pgf@anchor@pbsshape@in\endcsname{%
				\noexpand\pbsshape@port[0]{0}
			}%
		}{}%
		\pgfutil@ifundefined{pgf@anchor@pbsshape@out0}{
			\expandafter\xdef\csname pgf@anchor@pbsshape@out0\endcsname{%
				\noexpand\pbsshape@port[0]{1}
			}%
		}{}%
		\pgfutil@ifundefined{pgf@anchor@pbsshape@out1}{
			\expandafter\xdef\csname pgf@anchor@pbsshape@out1\endcsname{%
				\noexpand\pbsshape@port[1]{1}
			}%
		}{}%
		\pgfutil@ifundefined{pgf@anchor@pbsshape@out}{
			\expandafter\xdef\csname pgf@anchor@pbsshape@out\endcsname{%
				\noexpand\pbsshape@port[0]{1}
			}%
		}{}%
		\pgfmathsetcount{\portcount}{0}
		\pgfmathloop%
		\ifnum\the\portcount<\nport
		\pgfutil@ifundefined{pgf@anchor@pbsshape@p\the\portcount}{
			\expandafter\xdef\csname pgf@anchor@pbsshape@p\the\portcount\endcsname{%
				\noexpand\pbsshape@port[\the\portcount]{2}
			}%
		}{}%
		\pgfmathaddtocount{\portcount}{1}	
		\repeatpgfmathloop
	}
}

\def\pbsshape@port[#1]#2{
	\northeast	

	\ifnum#2=0	
		\if\direction\direce
			\ifnum#1=0
				\pgf@x=-\pgf@x
				\pgf@ya= \pgf@y
				\pgfmathsetlength{\pgf@y}{\pgf@ya-0.5*\minheight}%
			\else
				\pgf@x=0\pgf@x
				\pgf@ya= \pgf@y
				\pgfmathsetlength{\pgf@y}{\pgf@ya-\minheight}%
			\fi
		\fi
		\if\direction\direcw
			\ifnum#1=0
				\pgf@x=\pgf@x
				\pgf@ya= \pgf@y
				\pgfmathsetlength{\pgf@y}{\pgf@ya-0.5*\minheight}%
			\else
				\pgf@x=0\pgf@x
				\pgf@ya= \pgf@y
				\pgfmathsetlength{\pgf@y}{\pgf@ya}%
			\fi
		\fi
		\if\direction\direcn
			\ifnum#1=0
				\pgf@y=-\pgf@y
				\pgf@xa=\pgf@x
				\pgfmathsetlength{\pgf@x}{\pgf@xa-0.5*\minwidth}%
			\else
				\pgf@y=0\pgf@y
				\pgf@xa=\pgf@x
				\pgfmathsetlength{\pgf@x}{\pgf@xa-0*\minwidth}%
			\fi
		\fi
		\if\direction\direcs
			\ifnum#1=0
				\pgf@y=\pgf@y
				\pgf@xa= \pgf@x
				\pgfmathsetlength{\pgf@x}{\pgf@xa-0.5*\minwidth}%
			\else
				\pgf@y=0\pgf@y
				\pgf@xa= \pgf@x
				\pgfmathsetlength{\pgf@x}{\pgf@xa-1*\minwidth}%
			\fi
		\fi
	\else	
		\if\direction\direce
			\ifnum#1=0
				\pgf@x=\pgf@x
				\pgf@ya= \pgf@y
				\pgfmathsetlength{\pgf@y}{\pgf@ya-0.5*\minheight}%
			\else
				\pgf@x=0\pgf@x
				\pgf@ya= \pgf@y
				\pgfmathsetlength{\pgf@y}{\pgf@ya}%
			\fi
		\fi
		\if\direction\direcw
			\ifnum#1=0
				\pgf@x=-\pgf@x
				\pgf@ya= \pgf@y
				\pgfmathsetlength{\pgf@y}{\pgf@ya-0.5*\minheight}%
			\else
				\pgf@x=0\pgf@x
				\pgf@ya= \pgf@y
				\pgfmathsetlength{\pgf@y}{\pgf@ya-\minheight}%
			\fi
		\fi
		\if\direction\direcn
			\ifnum#1=0
				\pgf@y=\pgf@y
				\pgf@xa= \pgf@x
				\pgfmathsetlength{\pgf@x}{\pgf@xa-0.5*\minwidth}%
			\else
				\pgf@y=0\pgf@y
				\pgf@xa= \pgf@x
				\pgfmathsetlength{\pgf@x}{\pgf@xa-1*\minwidth}%
			\fi
		\fi
		\if\direction\direcs
			\ifnum#1=0
				\pgf@y=-\pgf@y
				\pgf@xa= \pgf@x
				\pgfmathsetlength{\pgf@x}{\pgf@xa-0.5*\minwidth}%
			\else
				\pgf@y=0\pgf@y
				\pgf@xa= \pgf@x
				\pgfmathsetlength{\pgf@x}{\pgf@xa-0*\minwidth}%
			\fi
		\fi
	\fi

	\ifnum#2=2	%
		\northeast \pgf@xa=\pgf@x \pgf@ya=\pgf@y
		\southwest \pgf@xb=\pgf@x \pgf@yb=\pgf@y

		\pgf@xc=\pgf@xa \advance\pgf@xc by -\pgf@xb	
		\pgfmathsetlength{\pgf@x}{\pgf@xa-(#1 + 0.5)*(\pgf@xc/\nport)}%
		\pgf@yc=\pgf@ya \advance\pgf@yc by -\pgf@yb	
		\pgfmathsetlength{\pgf@y}{\pgf@ya-(#1 + 0.5)*(\pgf@yc/\nport)}%

		\if\direction\direce
			\pgf@x=-\pgf@x
		\fi
		\if\direction\direcw
			\pgf@x=-\pgf@x
		\fi
	\fi
}

\pgfaddtoshape{pbsshape}{
	\anchorlet{c}{center}
	\anchorlet{n}{north}
	\anchorlet{e}{east}
	\anchorlet{s}{south}
	\anchorlet{w}{west}
	\anchorlet{se}{south east}
	\anchorlet{sw}{south west}
	\anchorlet{ne}{north east}
	\anchorlet{nw}{north west}
	\anchorlet{wsw}{west south west}
	\anchorlet{wnw}{west north west}
	\anchorlet{ene}{east north east}
	\anchorlet{ese}{east south east}
	\anchorlet{nnw}{north north west}
	\anchorlet{nne}{north north east}
	\anchorlet{ssw}{south south west}
	\anchorlet{sse}{south south east}
}

\pgfmathsetmacro{\WSSSINEHEIGHT}{0.06}

\tikzset{
	/tikz/pbskeys/.cd,
	size/.initial=0.5,
	color/.initial=O,
	direction/.initial=e,
	linestyle/.initial={linestyle},
	nport/.initial=1,
	fillgradient/.initial=O,
	/tikz/pbs/.code={
			\pgfqkeys{/tikz/pbskeys}{#1}%
			\tikzset{/tikz/pbskeys/drawer/.expanded=%
					{\pgfkeysvalueof{/tikz/pbskeys/direction}}%
					{\pgfkeysvalueof{/tikz/pbskeys/size}}%
					{\pgfkeysvalueof{/tikz/pbskeys/color}}%
					{\pgfkeysvalueof{/tikz/pbskeys/linestyle}}%
					{\pgfkeysvalueof{/tikz/pbskeys/fillgradient}}%
			}
		},
	/tikz/pbskeys/drawer/.code n args={5}{%
			\tikzset{
				pbsshape,
				minimum height=#2*\NODESIZE,
				minimum width=#2*\NODESIZE,
				#3,
				#4,
				draw,
				append after command={
						\pgfextra{\let\bdr=\tikzlastnode%
							\node[#5, fit=(\bdr.nw)(\bdr.se)] (boxgradient){};

							\if#1e
								\draw[#3, ---] ($(\bdr.nw)!.01!(\bdr.se)$) to ($(\bdr.se)!.01!(\bdr.nw)$);
							\fi
							\if#1w
								\draw[#3, ---] ($(\bdr.nw)!.01!(\bdr.se)$) to ($(\bdr.se)!.01!(\bdr.nw)$);
							\fi
							\if#1n
								\draw[#3, ---] ($(\bdr.ne)!.01!(\bdr.sw)$) to ($(\bdr.sw)!.01!(\bdr.ne)$);
							\fi
							\if#1s
								\draw[#3, ---] ($(\bdr.ne)!.01!(\bdr.sw)$) to ($(\bdr.sw)!.01!(\bdr.ne)$);
							\fi

						}
					}
			}
		},
}

\ifdefined\fontchoice
\else
	\def\fontchoice{times} 
\fi

\usepackage{pdftexcmds}

\ifnum\pdf@strcmp{\fontchoice}{firasans}=0 %
	\usepackage[book]{FiraSans} 
	\usepackage[T1]{fontenc}
	
	\tikzset{every picture/.style={/utils/exec={\sffamily}}}
\fi

\ifnum\pdf@strcmp{\fontchoice}{times}=0 %
	\usepackage{times}
\fi

\ifnum\pdf@strcmp{\fontchoice}{timesnewroman}=0 %
	\usepackage{mathptmx}
	\usepackage[T1]{fontenc}
\fi

\ifnum\pdf@strcmp{\fontchoice}{helvetica}=0 %
	\usepackage[scaled]{helvet}
	
	\usepackage[T1]{fontenc}
\fi

\makeatother
\pgfplotscreateplotcyclelist{foo}{
        {C1,mark=*},
        {C2,mark=square*},
        {C3, mark = triangle*},
        {C4,mark=+},
        {C5, mark = diamond*},
        {C6, mark = x}
      }
\newcommand{\SetCapsType}{normalcaps}
\usepackage[acronym,nomain]{glossaries}
\glsdisablehyper
\usepackage{xspace}  
\usepackage[shortcuts]{extdash}
\usepackage{listofitems,pgffor}    
\usepackage{siunitx}
\usepackage{xstring}    
\usepackage{silence}
\usepackage{xparse}

\setsepchar{;}  

\ifdefined\silencecommonwarnings
\else
	\def\silencecommonwarnings{true} 
\fi

\ifbool{\silencecommonwarnings}{%
    \WarningFilter{ECOtools}{Cannot define: DH}%
    \WarningFilter{ECOtools}{Cannot define: PAM}%
    \WarningFilter{ECOtools}{Cannot define: QAM}%
    \WarningFilter{ECOtools}{Cannot define: SI}%
    \WarningFilter{ECOtools}{Cannot define: PV}%
    \WarningFilter{ECOtools}{Cannot define: LP}%
    \WarningFilter{ECOtools}{Cannot define: RN}%
    \WarningFilter{ECOtools}{Cannot define: uLP}%
    \WarningFilter{ECOtools}{Redefining DH}%
    }{}

\makeatletter
\providecommand{\SetCapsType}{smallcaps}

\long\def\@scTrue{smallcaps}
\long\def\@scFalse{normalcaps}
\newcommand{\acroSCaps}[1]{%
    \ifx\SetCapsType\@scTrue 
        \textsc{#1}%
    \else
        \MakeUppercase{#1}%
    \fi
}
\makeatother

\usepackage{scalerel}
\makeatletter
\newcommand\scslash{%
\ifx\SetCapsType\@scTrue 
    \protect\stretchrel*{$/$}{\textsc{e}}
\else
    /
\fi
} 
\makeatother 

\makeatletter
\@ifpackageloaded{babel}{%
    \newcommand{\usuk}[2]{%
        \iflanguage{USenglish}{#1}{#2}%
    }%
}{%
    \newcommand{\usuk}[2]{%
        #1%
    }%
}%

\newcommand{\langcheck}[2]{
    \@ifpackageloaded{babel}{%
        \iflanguage{USenglish}{#1}{#2}%
    }{%
        #1%
    }%
}

\makeatother

\newcommand{\short}[1]{%
    \glsentrytext{#1}\xspace%
}
\newcommand{\shortfakeplural}[1]{%
    \glsentrytext{#1}s\xspace%
}
\newcommand{\Short}[1]{%
    \Glsentrytext{#1}\xspace%
}
\newcommand{\normal}[1]{%
    \gls{#1}\xspace%
}
\newcommand{\longacr}[1]{%
    \acrlong{#1}\xspace%
}
\newcommand{\plural}[1]{%
    \glspl{#1}\xspace%
}
\newcommand{\full}[1]{%
    \acrfull{#1}\xspace%
}
\newcommand{\fullplural}[1]{%
    \acrfullpl{#1}\xspace%
}
\newcommand{\Normal}[1]{%
    \Gls{#1}\xspace%
}
\newcommand{\Plural}[1]{%
    \Glspl{#1}\xspace%
}
\newcommand{\Full}[1]{%
    \Acrfull{#1}\xspace%
}
\newcommand{\Fullplural}[1]{%
    \Acrfullpl{#1}\xspace%
} 

\newcommand{\texpdfif}[2]{%
    \ifcsname texorpdfstring\endcsname%
        \texorpdfstring{#1{#2}}{#2\xspace}%
    \else%
        #1{#2}%
    \fi%
}

\newcommand{\checkanddefine}[3]{%
	\ifcsname #1\endcsname%
        \PackageWarning{ECOtools}{Cannot define: #1 already defined, trying to define g#1 instead.}%
        \ifcsname g#1\endcsname%
            \PackageWarning{ECOtools}{Cannot define: g#1 also already defined.}%
    	\else%
        	\expandafter\newcommand\csname g#1\endcsname{%
        	    \texpdfif{#2}{#3}%
    	    }%
        \fi%
	\else%
    	\expandafter\newcommand\csname #1\endcsname{%
    	    \texpdfif{#2}{#3}%
	    }%
    \fi%
}

\newcommand{\redefine}[3]{%
    \PackageWarning{ECOtools}{Redefining #1}%
	\expandafter\renewcommand\csname #1\endcsname{%
	    \texpdfif{#2}{#3}%
    }%
}

\newcommand{\nAcronym}[4][]{%
	\newacronym[#1]{#2}{#3}{#4}%
	\checkanddefine{s#2}{\short}{#2}%
    \checkanddefine{s#2s}{\shortfakeplural}{#2}%
	\checkanddefine{#2}{\normal}{#2}%
	\checkanddefine{l#2}{\longacr}{#2}%
	\checkanddefine{#2s}{\plural}{#2}%
	\checkanddefine{f#2}{\full}{#2}%
	\checkanddefine{f#2s}{\fullplural}{#2}%
	\checkanddefine{su#2}{\Short}{#2}%
	\checkanddefine{u#2}{\Normal}{#2}%
	\checkanddefine{u#2s}{\Plural}{#2}%
	\checkanddefine{fu#2}{\Full}{#2}%
	\checkanddefine{fu#2s}{\Fullplural}{#2}%
	\IfStrEq{#2}{DH}{
	    \redefine{#2}{\normal}{#2}%
	    }{}%
}%

\NewDocumentCommand\qam{g}{%
    \IfNoValueTF{#1}{%
        \texpdfif{\gls}{QAM}\xspace%
        }{%
        \StrLen{#1}[\stringlength]%
        \ifnum\stringlength=0%
            \texpdfif{\gls}{QAM}\xspace%
        \else%
            {\qamlisthelper{#1}}%
        \fi%
        }%
}

\let\QAM\qam

\DeclareRobustCommand\qamlisthelper[1]{%
    \readlist*\args{#1}%
    \acroSCaps{\args[1]\=/}%
    \ifnum\argslen = 2%
        { and \acroSCaps{\args[2]}\=/}%
    \fi%
    \ifnum\argslen > 2%
        \foreach \n in {2,...,\argslen}{%
            \ifnum\n = \argslen%
                {, and }%
            \else 
                {, }%
            \fi%
            {\acroSCaps{\args[\n]}\=/}%
        }%
    \fi%
    \ifglsused{QAM}%
        {}%
        {ary }%
    \texpdfif{\gls}{QAM}%
}%

\NewDocumentCommand\pam{g}{%
    \IfNoValueTF{#1}{%
        \texpdfif{\gls}{PAM}\xspace%
        }{%
        \StrLen{#1}[\stringlength]%
        \ifnum\stringlength=0%
            \texpdfif{\gls}{PAM}\xspace%
        \else%
            {\pamlisthelper{#1}}%
        \fi%
        }%
}

\DeclareRobustCommand\pamlisthelper[1]{%
    \readlist*\args{#1}%
    \ifglsused{PAM}{%
        \texpdfif{\gls}{PAM}%
        \acroSCaps{\=/\args[1]}%
        \ifnum\argslen = 2%
            { and \=/\acroSCaps{\args[2]}}%
        \fi%
        \ifnum\argslen > 2%
            \foreach \n in {2,...,\argslen}{%
                \ifnum\n = \argslen%
                    {, and }%
                \else%
                    {, }%
                \fi%
                {\=/\acroSCaps{\args[\n]}}%
            }%
        \fi%
    }{%
        \acroSCaps{\args[1]\=/}%
        \ifnum\argslen = 2%
            { and \acroSCaps{\args[2]}\=/}%
        \fi%
        \ifnum\argslen > 2%
            \foreach \n in {2,...,\argslen}{%
                \ifnum\n = \argslen%
                    {, and }%
                \else%
                    {, }%
                \fi
                {\acroSCaps{\args[\n]}\=/}%
            }%
        \fi%
        {ary }%
        \texpdfif{\gls}{PAM}%
    }%
}%

\NewDocumentCommand\lp{g}{%
    \IfNoValueTF{#1}{%
        \texpdfif{\normal}{LP}%
        }{%
        \StrLen{#1}[\stringlength]%
        \ifnum\stringlength=0%
            \texpdfif{\normal}{LP}%
        \else%
            \ifglsused{LP}{}{\texpdfif{\normal}{LP}\xspace}%
            \lplisthelper[lp]{#1}%
        \fi%
        }%
}

\NewDocumentCommand\ulp{g}{%
    \IfNoValueTF{#1}{%
        \texpdfif{\Normal}{LP}\xspace%
        }{%
        \StrLen{#1}[\stringlength]%
        \ifnum\stringlength=0%
            \texpdfif{\Normal}{LP}\xspace%
        \else%
            \ifglsused{LP}{%
                \lplisthelper[Lp]{#1}%
            }{%
                \texpdfif{\Normal}{LP}\xspace\lplisthelper[lp]{#1}%
            }%
        \fi%
        }%
}
\DeclareRobustCommand\lplisthelper[2][lp]{%
    \readlist*\args{#2}%
    \foreach \n in {1,...,\argslen}{%
        \ifnum \n > 1%
            \ifnum \argslen > 2%
                {, }%
            \else%
                { }%
            \fi%
        \fi%
        \ifnum \n = \argslen%
            \ifnum \argslen > 1%
                {and }%
            \fi%
        \fi%
        \ifnum \n = 1%
            {\acroSCaps{#1}}
        \else%
            {\acroSCaps{\MakeLowercase{#1}}}%
        \fi%
        {\textsubscript{\StrSplit{\args[\n]}{2}{\csA}{\csB}\acroSCaps{\csA}\csB}}
    }%
}%

\nAcronym{128SPQAM}{\acroSCaps{128-sp-16-qam}}{128-ary set-partitioning \QAM{16}}

\nAcronym{2A8PSK}{\acroSCaps{2a8psk}}{2-ary amplitude 8-ary phaseshift keying}

\nAcronym{3CCMCF}{\acroSCaps{3cc-mcf}}{3-core coupled-core multi-core fiber}

\nAcronym{4D}{\acroSCaps{4d}}{four-dimensional}
\nAcronym{4D64PRS}{\acroSCaps{4d-64prs}}{\usuk{four-dimensional 64-ary polarization-ring-switching}{four-dimensional 64-ary polarisation-ring-switching}}
\nAcronym{4DOS128}{\acroSCaps{4d-os128}}{four-dimensional orthant-symmetric 128-ary modulation format}

\nAcronym{5B4D2A8PSK}{\acroSCaps{5b4d-2a8psk}}{5-bit four-dimensional two-amplitude 8-ary phase-shift keying}

\nAcronym{6B4D2A8PSK}{\acroSCaps{6b4d-2a8psk}}{6-bit four-dimensional two-amplitude 8-ary phase-shift keying}

\nAcronym{7B4D2A8PSK}{\acroSCaps{7b4d-2a8psk}}{7-bit four-dimensional two-amplitude 8-ary phase-shift keying}

\nAcronym{8D}{\acroSCaps{8d}}{eight-dimensional}\nAcronym{8D2048PRS}{\acroSCaps{8d-2048prs}}{eight-dimensional 2048-ary polarization-ring-switching}
\nAcronym{8D2048PRST1}{\acroSCaps{8d-2048prs-t1}}{eight-dimensional 2048-ary polarization-ring-switching type 1}
\nAcronym{8D2048PRST2}{\acroSCaps{8d-2048prs-t2}}{eight-dimensional 2048-ary polarization-ring-switching type 2}
\nAcronym{8DAPSK}{\acroSCaps{8d-apsk}}{eight-dimensional amplitude-phase-shift keying}

\nAcronym{ABC}{\acroSCaps{abc}}{automatic bias control}
\nAcronym{AC}{\acroSCaps{ac}}{alternating current}
\nAcronym{ADC}{\acroSCaps{adc}}{\usuk{analog-to-digital converter}{analogue-to-digital converter}}
\nAcronym{AGC}{\acroSCaps{agc}}{automatic gain control}
\nAcronym{AIR}{\acroSCaps{air}}{achievable information rate}
\nAcronym{AMZI}{\acroSCaps{amzi}}{asymmetric Mach–Zehnder interferometer}
\nAcronym{AO}{\acroSCaps{ao}}{adaptive optics}
\nAcronym{AOM}{\acroSCaps{aom}}{acousto-optic modulator}
\nAcronym{APD}{\acroSCaps{apd}}{avalanche photodiode}
\nAcronym{API}{\acroSCaps{api}}{application programming interface}
\nAcronym{AR}{\acroSCaps{ar}}{achievable rate}
\nAcronym{ARRWG}{\acroSCaps{a}rr\acroSCaps{wg}}{arrayed-waveguide grating}
\nAcronym{ASE}{\acroSCaps{ase}}{amplified spontaneous emission}
\nAcronym{ASK}{\acroSCaps{ask}}{amplitude-shift keying}
\nAcronym{ASIC}{\acroSCaps{asic}}{application-specific integrated circuit}
\nAcronym{ATS}{\acroSCaps{ats}}{alignment tracking sensor}
\nAcronym{AWG}{\acroSCaps{awg}}{arbitrary-waveform generator}
\nAcronym{AWGN}{\acroSCaps{awgn}}{additive white Gaussian noise}

\nAcronym{BBU}{\acroSCaps{bbu}}{baseband unit}
\nAcronym{BCH}{\acroSCaps{bch}}{Bose-Chaudhuri-Hocquenghem}
\nAcronym{BER}{\acroSCaps{ber}}{bit error rate}
\nAcronym{BERT}{\acroSCaps{bert}}{bit error rate tester}
\nAcronym{BICM}{\acroSCaps{bicm}}{bit-interleaved coded modulation}
\nAcronym{BMD}{\acroSCaps{bmd}}{bit-metric decoding}
\nAcronym{BPD}{\acroSCaps{bpd}}{balanced photo-diode}
\nAcronym{BPF}{\acroSCaps{bpf}}{bandpass filter}
\nAcronym{BPS}{\acroSCaps{bps}}{blind phase search}
\nAcronym{BPSK}{\acroSCaps{bpsk}}{binary phase-shift keying}
\nAcronym{BRGC}{\acroSCaps{brgc}}{binary reflected Gray code}
\nAcronym{BTB}{\acroSCaps{btb}}{back-to-back}

\nAcronym{CAGR}{\acroSCaps{cagr}}{compound annual growth rate}
\nAcronym{CCDM}{\acroSCaps{ccdm}}{constant composition distribution matching}
\langcheck{%
    \nAcronym{CCF}{\acroSCaps{ccf}}{coupled-core fiber}%
    }{%
    \nAcronym{CCF}{\acroSCaps{ccf}}{coupled-core fibre}%
}%
\nAcronym{CD}{\acroSCaps{cd}}{chromatic dispersion}
\nAcronym{CIR}{\acroSCaps{cir}}{channel impulse response}
\nAcronym{CMA}{\acroSCaps{cma}}{constant modulus algorithm}
\nAcronym{CMF}{\acroSCaps{cmf}}{core multiplicity factor}
\nAcronym{CMUX}{\acroSCaps{cmux}}{core multiplexer}
\nAcronym{COTS}{\acroSCaps{cots}}{commercial off-the-shelf}
\nAcronym{COW}{\acroSCaps{cow}}{coherent one-way}
\nAcronym{ChUT}{\acroSCaps{chut}}{channel under test}
\nAcronym[firstplural=channels under test (\acroSCaps{cut}s)]{CUT}{\acroSCaps{cut}}{channel under test}
\nAcronym{CRX}{\acroSCaps{crx}}{coherent receiver}
\nAcronym{CPE}{\acroSCaps{cpe}}{carrier phase estimation}
\nAcronym{CPU}{\acroSCaps{cpu}}{central processing unit}
\nAcronym{CSPR}{\acroSCaps{cspr}}{carrier-to-signal power ratio}
\nAcronym{CUDA}{\acroSCaps{cuda}}{compute unified device architecture}
\nAcronym{CVQKD}{\acroSCaps{cv-qkd}}{continuous-variable quantum key distribution}
\nAcronym{CW}{\acroSCaps{cw}}{continuous wave}
\nAcronym{CCD}{\acroSCaps{ccd}}{charge-coupled device}

\nAcronym{DA}{\acroSCaps{da}}{driver amplifier}
\nAcronym{DAC}{\acroSCaps{dac}}{\usuk{digital-to-analog converter}{digital-to-analogue converter}}
\nAcronym{DC}{\acroSCaps{dc}}{direct current}
\nAcronym{DBP}{\acroSCaps{dbp}}{digital backpropagation}
\nAcronym{DCF}{\acroSCaps{dcf}}{\usuk{dispersion-compensating fiber}{dispersion-compensating fibre}}
\langcheck{%
    \nAcronym{DCI}{\acroSCaps{dci}}{data center interconnect}
    }{%
    \nAcronym{DCI}{\acroSCaps{dci}}{data centre interconnect}
}%
\nAcronym{DDLMS}{\acroSCaps{dd-lms}}{decision-directed least mean square}
\nAcronym{DEMUX}{\acroSCaps{demux}}{de-multiplexer}
\nAcronym{DFA}{\acroSCaps{dfa}}{\usuk{doped fiber amplifier}{doped fibre amplifier}}
\nAcronym{DFB}{\acroSCaps{dfb}}{distributed feedback}
\nAcronym{DGD}{\acroSCaps{dgd}}{differential group delay}
\nAcronym{DH}{\acroSCaps{dh}}{digital holography}
\nAcronym{DM}{\acroSCaps{dm}}{distribution matcher}
\nAcronym{DMR}{\acroSCaps{dm}}{dichroic mirror}
\nAcronym{DMA}{\acroSCaps{dma}}{direct memory access}
\nAcronym{DMD}{\acroSCaps{dmd}}{differential mode delay}
\nAcronym{DMG}{\acroSCaps{dmg}}{differential modal gain}
\nAcronym{DMGD}{\acroSCaps{dmgd}}{differential mode group delay}
\nAcronym{DML}{\acroSCaps{dml}}{directly-modulated laser}
\nAcronym{DP}{\acroSCaps{dp}}{\usuk{dual-polarization}{dual-polarisation}}
\nAcronym{DPC}{\acroSCaps{dpc}}{digital pre-compensation}
\nAcronym{DPE}{\acroSCaps{dpe}}{digital pre-emphasis}
\nAcronym{DPIQ}{\acroSCaps{dp-iqm}}{\usuk{dual-polarization \acroSCaps{iq}-modulator}{dual-polarisation \acroSCaps{iq}-modulator}}
\nAcronym{DPLL}{\acroSCaps{dpll}}{digital phase-locked loop}
\nAcronym{DPS}{\acroSCaps{dps}}{differential phase shift}
\nAcronym{DQPSK}{\acroSCaps{dqpsk}}{differential quaternary phase-shift-keying}
\nAcronym{DRA}{\acroSCaps{dra}}{distributed Raman amplifier}
\nAcronym{DRE}{\acroSCaps{dre}}{digital resolution enhancer}
\nAcronym{DSB}{\acroSCaps{dsb}}{double-sideband}
\nAcronym{DSF}{\acroSCaps{dsf}}{\usuk{dispersion-shifted fiber}{dispersion-shifted fibre}} 
\nAcronym{DSO}{\acroSCaps{dso}}{digital sampling oscilloscope}
\nAcronym{DSP}{\acroSCaps{dsp}}{digital signal processing}
\nAcronym{DUT}{\acroSCaps{dut}}{device-under-test}

\nAcronym{DVQKD}{\acroSCaps{dv-qkd}}{discrete-variable quantum key distribution}

\nAcronym{DWDM}{\acroSCaps{dwdm}}{dense wavelength-division multiplexing}

\nAcronym{EAM}{\acroSCaps{eam}}{electro-absorption modulator}
\nAcronym{ECL}{\acroSCaps{ecl}}{external cavity laser}
\nAcronym{ED}{\acroSCaps{ed}}{Eucledian distance}
\nAcronym{EDF}{\acroSCaps{edf}}{\usuk{erbium-doped fiber}{erbium-doped fibre}}
\nAcronym{EDFA}{\acroSCaps{edfa}}{\usuk{erbium-doped fiber amplifier}{erbium-doped fibre amplifier}}
\nAcronym{ENOB}{\acroSCaps{enob}}{effective number of bits}
\nAcronym{ER}{\acroSCaps{er}}{extinction ratio}
\nAcronym{ESS}{\acroSCaps{ess}}{enumerative sphere shaping}

\langcheck{%
    \nAcronym{FBG}{\acroSCaps{fbg}}{fiber Bragg grating}%
    }{%
    \nAcronym{FBG}{\acroSCaps{fbg}}{fibre Bragg grating}%
}%
\nAcronym{FD}{\acroSCaps{fd}}{frequency domain}
\nAcronym{FDE}{\acroSCaps{fde}}{\usuk{frequency domain equalizer}{frequency domain equaliser}}
\nAcronym{FEC}{\acroSCaps{fec}}{forward error correction}
\nAcronym{FFE}{\acroSCaps{ffe}}{\usuk{feed-forward equalizer}{feed-forward equaliser}}
\nAcronym{FFT}{\acroSCaps{fft}}{fast Fourier transform}
\nAcronym{FIR}{\acroSCaps{fir}}{finite impulse response}
\nAcronym{FLOPS}{\acroSCaps{flops}}{floating point operations per second}
\nAcronym{FMEDF}{\acroSCaps{fm-edf}}{\usuk{few-mode erbium-doped fiber}{few-mode erbium-doped fibre}}
\nAcronym{FMEDFA}{\acroSCaps{fm-edfa}}{\usuk{few-mode erbium-doped fiber amplifier}{few-mode erbium-doped fibre amplifier}}
\langcheck{%
    \nAcronym{FMF}{\acroSCaps{fmf}}{few-mode fiber}%
    }{%
    \nAcronym{FMF}{\acroSCaps{fmf}}{few-mode fibre}%
}%
\nAcronym[plural=FM-MCF, firstplural=\usuk{few-mode multi-core fibers}{few-mode multi-core fibres}]{FMMCF}{\acroSCaps{fm-mcf}}{\usuk{few-mode multi-core fiber}{few-mode multi-core fibre}}
\langcheck{%
    \nAcronym{FMPBGF}{\acroSCaps{fm-pbgf}}{few-mode photonic bandgap fiber}%
    }{%
    \nAcronym{FMPBGF}{\acroSCaps{fm-pbgf}}{few-mode photonic bandgap fibre}%
}%
\nAcronym{FOV}{\acroSCaps{fov}}{field of view}
\nAcronym{FPGA}{\acroSCaps{fpga}}{field-programmable gate array}
\nAcronym{FWM}{\acroSCaps{fwm}}{four-wave mixing}
\nAcronym{FSO}{\acroSCaps{fso}}{free-space optical}
\nAcronym{FUT}{\acroSCaps{fut}}{\usuk{fiber under test}{fibre under test}}

\nAcronym{GD}{\acroSCaps{gd}}{group delay}
\nAcronym{GI}{\acroSCaps{gi}}{graded-index}
\nAcronym{GFF}{\acroSCaps{gff}}{gain flattening filter}
\nAcronym{GIFMF}{\acroSCaps{gi-fmf}}{\usuk{graded-index few-mode fiber}{graded-index few-mode fibre}}
\nAcronym{GIMMF}{\acroSCaps{gi-mmf}}{\usuk{graded-index multi-mode fiber}{graded-index multi-mode fibre}}
\nAcronym{GMI}{\acroSCaps{gmi}}{\usuk{generalized mutual information}{generalised mutual information}}
\nAcronym{GNSE}{\acroSCaps{gnse}}{generalized nonlinear Schr\"{o}dinger equation}
\nAcronym{GPU}{\acroSCaps{gpu}}{graphics processing unit}
\nAcronym{GS}{\acroSCaps{gs}}{geometric shaping}
\nAcronym{GV}{\acroSCaps{gv}}{group velocity}
\nAcronym{GVD}{\acroSCaps{gvd}}{group velocity dispersion}
\nAcronym{GPIO}{\acroSCaps{gpio}}{general purpose input output}
\nAcronym{GUI}{\acroSCaps{gui}}{graphical user interface}

\langcheck{%
    \nAcronym{HCF}{\acroSCaps{hcf}}{hollow-core fiber}
    }{%
    \nAcronym{HCF}{\acroSCaps{hcf}}{hollow-core fibre}
}%
\nAcronym{HDFEC}{\acroSCaps{hd-fec}}{hard-decision forward error correction}
\nAcronym{HG}{\acroSCaps{hg}}{Hermite-Gaussian}
\nAcronym{HOM}{\acroSCaps{hom}}{higher-order modes}
\nAcronym{HV}{\acroSCaps{hv}}{Hufnagel-Valley}
\nAcronym{HAP}{\acroSCaps{hap}}{Hufnagel-Andrew-Phillips}

\nAcronym{ICS}{\acroSCaps{ics}}{inter-core skew}
\nAcronym{ICXT}{\acroSCaps{ic-xt}}{inter-core cross-talk}
\nAcronym[plural=IL, firstplural=insertion losses (\acroSCaps{il})]{IL}{\acroSCaps{il}}{insertion loss}
\nAcronym{IFFT}{\acroSCaps{ifft}}{inverse fast Fourier transform}
\nAcronym{IIR}{\acroSCaps{iir}}{intensity impulse response}
\nAcronym{IM}{\acroSCaps{im}}{intensity modulator}
\nAcronym{IMDD}{\acroSCaps{im}\scslash \acroSCaps{dd}}{intensity-modulation direct-detection}
\nAcronym{IQM}{\acroSCaps{iqm}}{in-phase and quadrature modulator}
\nAcronym{ISI}{\acroSCaps{isi}}{inter-symbol interference}
\nAcronym{IP}{\acroSCaps{ip}}{intellectual property}

\nAcronym{JGN}{\acroSCaps{jgn}}{Japan Gigabit Network}

\nAcronym{KK}{\acroSCaps{kk}}{Kramers-Kronig}
\nAcronym{KIT}{\acroSCaps{kit}}{Karlsruhe Institute of Technology}

\nAcronym{LCOS}{\acroSCaps{LCoS}}{liquid crystal on silicon}
\nAcronym{LDPC}{\acroSCaps{ldpc}}{low-density parity-check}
\nAcronym{LEAF}{\acroSCaps{leaf}}{\usuk{large effective area fiber}{large effective area fibre}}
\nAcronym{LFSR}{\acroSCaps{lfsr}}{linear-feedback shift register}
\nAcronym{LG}{\acroSCaps{lg}}{Laguerre-Gaussian}
\nAcronym{LMS}{\acroSCaps{lms}}{least mean square}
\nAcronym{LLR}{\acroSCaps{llr}}{log-likelihood ratio}
\nAcronym{LO}{\acroSCaps{lo}}{local oscillator}
\nAcronym{LP}{\acroSCaps{lp}}{\usuk{linearly polarized}{linearly polarised}}
\nAcronym{LSPS}{\acroSCaps{lsps}}{\usuk{loop-synchronized polarization scrambler}{loop-synchronised polarisation scrambler}}
\nAcronym{LUT}{\acroSCaps{lut}}{lookup table}

\nAcronym{MVM}{\acroSCaps{mvm}}{matrix-vector multiplication}
\nAcronym{MB}{\acroSCaps{mb}}{Maxwell-Bolzmann}
\langcheck{%
    \nAcronym{MCF}{\acroSCaps{mcf}}{multi-core fiber}%
    }{%
    \nAcronym{MCF}{\acroSCaps{mcf}}{multi-core fibre}%
}%
\nAcronym{MDG}{\acroSCaps{mdg}}{mode dependent gain}
\nAcronym[firstplural=mode-dependent losses (\acroSCaps{mdl})]{MDL}{\acroSCaps{mdl}}{mode-dependent loss}
\nAcronym{MDM}{\acroSCaps{mdm}}{mode-division multiplexing}
\nAcronym{MEMS}{\acroSCaps{mems}}{micro-electro-mechanical systems}
\nAcronym{MF}{\acroSCaps{mf}}{matched filter}
\nAcronym{MFD}{\acroSCaps{mfd}}{mode field diameter}
\nAcronym{MI}{\acroSCaps{mi}}{mutual information}
\nAcronym{MIMO}{\acroSCaps{mimo}}{multiple-input multiple-output}
\nAcronym{ML}{\acroSCaps{ml}}{machine learning}
\nAcronym{MMA}{\acroSCaps{mma}}{multi-modulus algorithm}
\nAcronym{MMEDF}{\acroSCaps{mmedf}}{\usuk{multi-mode erbium-doped fiber}{multi-mode erbium-doped fibre}}
\nAcronym{MMEDFA}{\acroSCaps{mmedfa}}{\usuk{multi-mode erbium-doped fiber amplifier}{multi-mode erbium-doped fibre amplifier}}
\nAcronym{MMF}{\acroSCaps{mmf}}{\usuk{multi-mode fiber}{multi-mode fibre}}
\nAcronym{MMSE}{\acroSCaps{mmse}}{minimum mean squared error}
\nAcronym{MP}{\acroSCaps{mp}}{minimum phase}
\nAcronym{MPLC}{\acroSCaps{mplc}}{multi-plane light converter}
\nAcronym{MRC}{\acroSCaps{mrc}}{maximum ratio combining}
\nAcronym{MSE}{\acroSCaps{mse}}{mean squared error}
\nAcronym{MUX}{\acroSCaps{mux}}{multiplexer}
\nAcronym{MZM}{\acroSCaps{mzm}}{Mach-Zehnder modulator}
\nAcronym{MZI}{\acroSCaps{mzi}}{Mach-Zehnder interferometer}

\nAcronym{NA}{\acroSCaps{na}}{numerical aperture}
\langcheck{%
    \nAcronym{NANF}{\acroSCaps{nanf}}{nested antiresonant nodeless fiber}%
    }{%
    \nAcronym{NANF}{\acroSCaps{nanf}}{nested antiresonant nodeless fibre}%
}%
\nAcronym{NF}{\acroSCaps{nf}}{noise figure}
\nAcronym{NGMI}{\acroSCaps{ngmi}}{\usuk{normalized generalized mutual information}{normalised generalised mutual information}}
\nAcronym{NLSE}{\acroSCaps{nlse}}{nonlinear Schr\"{o}ding equation}
\nAcronym{NN}{\acroSCaps{nn}}{neural network}
\nAcronym{NIC}{\acroSCaps{nic}}{network interface card}
\nAcronym{NICT}{\acroSCaps{nict}}{National Institute of Information and Communications Technology}
\nAcronym{NIR}{\acroSCaps{nir}}{near-infrared}
\nAcronym{NISTSTS}{\acroSCaps{nist-sts}}{National Insitute of Standards and Technology: Statistical Test Suite}
\nAcronym{NRZ}{\acroSCaps{nrz}}{non-return-to-zero}
\nAcronym{NZDSF}{\acroSCaps{nz-dsf}}{\usuk{non-zero dispersion-shifted fiber}{non-zero dispersion-shifted fibre}} 

\nAcronym{OAM}{\acroSCaps{oam}}{orbital angular momentum}
\nAcronym{OBTB}{\acroSCaps{obtb}}{optical back-to-back}
\nAcronym{OCT}{\acroSCaps{oct}}{outer cladding thickness}
\nAcronym{ODE}{\acroSCaps{ode}}{ordinary differential equation}
\nAcronym{ODL}{\acroSCaps{odl}}{optical delay line}
\nAcronym{OEO}{\acroSCaps{oeo}}{optical-electrical-optical}
\nAcronym{OFC}{\acroSCaps{ofc}}{Optical Fiber Communications Conference}
\nAcronym{OFDR}{\acroSCaps{ofdr}}{optical frequency-domain reflectometer} 
\nAcronym{OFDM}{\acroSCaps{ofdm}}{orthogonal frequency division multiplexing}
\nAcronym{OH}{\acroSCaps{oh}}{overhead}
\nAcronym{OMFT}{\acroSCaps{omft}}{optical multi-format transmitter}
\nAcronym{OOK}{\acroSCaps{ook}}{on-off keying}
\nAcronym{OP}{\acroSCaps{op}}{optical processor}
\nAcronym{OPLL}{\acroSCaps{opll}}{optical phase-locked loop}
\nAcronym{OSA}{\acroSCaps{osa}}{\usuk{optical spectrum analyzer}{optical spectrum analyser}}
\nAcronym{OSNR}{\acroSCaps{osnr}}{optical signal-to-noise ratio}
\nAcronym{OTDR}{\acroSCaps{otdr}}{optical time-domain reflectometer}
\nAcronym{OTF}{\acroSCaps{otf}}{optical tunable filter}
\langcheck{%
    \nAcronym{OVNA}{\acroSCaps{ovna}}{optical vector network analyzer}%
    }{%
    \nAcronym{OVNA}{\acroSCaps{ovna}}{optical vector network analyser}%
}%
\nAcronym{OTG}{\acroSCaps{otg}}{optical turbulence generator}

\nAcronym{PAM}{\acroSCaps{pam}}{pulse-amplitude modulation}
\nAcronym{PAS}{\acroSCaps{pas}}{probabilistic amplitude shaping}
\nAcronym{PAPR}{\acroSCaps{papr}}{peak-to-average power ratio}
\nAcronym{PBC}{\acroSCaps{pbc}}{\usuk{polarization beam combiner}{polarisation beam combiner}}
\langcheck{%
    \nAcronym{PBGF}{\acroSCaps{pbgf}}{photonic bandgap fiber}%
    }{%
    \nAcronym{PBGF}{\acroSCaps{pbgf}}{photonic bandgap fibre}%
}%
\nAcronym{PBS}{\acroSCaps{pbs}}{polarization beam splitter}
\nAcronym{PC}{\acroSCaps{pc}}{physical contact}
\nAcronym{PCVD}{\acroSCaps{pcvd}}{plasma chemical vapor depostion}
\nAcronym{PCG}{\acroSCaps{pcg64}}{64-bit permuted congruential generator}
\nAcronym{PD}{\acroSCaps{pd}}{photodiode}
\nAcronym{PDF}{\acroSCaps{pdf}}{probability density function}
\langcheck{%
    \nAcronym{PDL}{\acroSCaps{pdl}}{polarization-dependent loss}
    }{%
    \nAcronym{PDL}{\acroSCaps{pdl}}{polarisation-dependent loss}
}%
\nAcronym{PDM}{\acroSCaps{pdm}}{\usuk{polarization-division multiplexing}{polarisation-division multiplexing}}
\langcheck{%
    \nAcronym{PER}{\acroSCaps{per}}{polarization extinction ratio}%
    }{%
    \nAcronym{PER}{\acroSCaps{per}}{polarisation extinction ratio}%
}%
\nAcronym{PIC}{\acroSCaps{pic}}{photonic integrated circuit}
\nAcronym{PL}{\acroSCaps{pl}}{photonic lantern}
\nAcronym{PM}{\acroSCaps{pm}}{polarization-multiplexed}
\nAcronym{PMBPSK}{\acroSCaps{pm-bpsk}}{polarization-multiplexed binary phase-shift keying}
\nAcronym{PMQPSK}{\acroSCaps{pm-qpsk}}{polarization-multiplexed quaternary phase-shift keying}
\nAcronym{PM8QAM}{\acroSCaps{pm-8qam}}{polarization-multiplexed 8-ary quadrature amplitude modulation}
\nAcronym{PMD}{\acroSCaps{pmd}}{\usuk{polarization mode dispersion}{polarisation mode dispersion}}
\langcheck{%
    \nAcronym{PMF}{\acroSCaps{pmf}}{polarization-maintaining fiber}%
    }{%
    \nAcronym{PMF}{\acroSCaps{pmf}}{polarization-maintaining fibre}%
}%
\nAcronym{PMP}{\acroSCaps{pmp}}{phase-matching point}
\nAcronym{PNOB}{\acroSCaps{pnob}}{physical number of bits}
\nAcronym{PON}{\acroSCaps{pon}}{passive-optical network}
\nAcronym{PRBS}{\acroSCaps{prbs}}{pseudorandom bit sequence}
\nAcronym{PRNG}{\acroSCaps{PRNG}}{pseudo random number generator}
\nAcronym{PROFA}{\acroSCaps{profa}}{\usuk{pitch reducing optical fiber array}{pitch reducing optical fibre array}}
\nAcronym{PPM}{\acroSCaps{ppm}}{pulse-position modulation}
\nAcronym{PS}{\acroSCaps{ps}}{probabilistic shaping}
\langcheck{%
    \nAcronym{PSCF}{\acroSCaps{pscf}}{pure silica core fiber}%
    }{%
    \nAcronym{PSCF}{\acroSCaps{pscf}}{pure silica core fibre}%
}%
\nAcronym{PSD}{\acroSCaps{psd}}{power spectral density}
\nAcronym{PSF}{\acroSCaps{psf}}{point spread function}
\nAcronym{PSK}{\acroSCaps{psk}}{phase-shift keying}
\nAcronym{PSP}{\acroSCaps{psp}}{\usuk{principal states of polarization}{principal states of polarisation}}
\nAcronym{PSW}{\acroSCaps{psw}}{\usuk{polarization switch}{polarisation switch}}
\nAcronym{PID}{\acroSCaps{pid}}{proportional–integral–derivative}
\nAcronym{PV}{\acroSCaps{pv}}{process value}

\nAcronym{QAM}{\acroSCaps{qam}}{quadrature amplitude modulation}
\nAcronym{QBER}{\acroSCaps{qber}}{quantum bit error rate}
\nAcronym{QKD}{\acroSCaps{qkd}}{quantum key distribution}
\nAcronym{QPSK}{\acroSCaps{qpsk}}{quadrature phase-shift keying}
\nAcronym{QRNG}{\acroSCaps{qrng}}{quantum random number generator}
\nAcronym{QSM}{\acroSCaps{qsm}}{quasi-single-mode}

\nAcronym{RAM}{\acroSCaps{ram}}{random-access memory}
\nAcronym{RC}{\acroSCaps{rc}}{raised cosine}
\nAcronym{RCMF}{\acroSCaps{rcmf}}{relative core multiplicity factor}
\nAcronym{RCMCF}{\acroSCaps{rc-mcf}}{\usuk{randomly-coupled multi-core fiber}{randomly-coupled multi-core fibre}}
\nAcronym{RF}{\acroSCaps{rf}}{radio frequency}
\nAcronym{RFSoC}{\acroSCaps{rfsoc}}{radio frequency system-on-chip}
\nAcronym{RI}{\acroSCaps{ri}}{refractive index}
\nAcronym{RLS}{\acroSCaps{rls}}{recursive least squares}
\nAcronym{RRC}{\acroSCaps{rrc}}{root-raised-cosine}
\nAcronym{ROADM}{\acroSCaps{roadm}}{reconfigurable optical add-drop multiplexer}
\nAcronym{ROI}{\acroSCaps{roi}}{region of interest} 
\nAcronym{RTO}{\acroSCaps{rto}}{real-time oscilloscope} 
\nAcronym{RZDBPSK}{\acroSCaps{rz-dbpsk}}{return-to-zero differential binary phase-shift keying} 
\nAcronym{RZDQPSK}{\acroSCaps{rz-dqpsk}}{return-to-zero differential quaternary phase-shift keying}

\nAcronym{S2}{\acroSCaps{S\textsuperscript{2}}}{spatially and spectrally resolved}
\nAcronym{SA}{\acroSCaps{sa}}{simulated annealing}
\nAcronym{SamPerSym}{\acroSCaps{sps}}{samples per symbol}
\nAcronym{SBS}{\acroSCaps{sbs}}{stimulated Brillouin scattering}
\nAcronym{SCM}{\acroSCaps{scm}}{subcarrier multiplexing}
\nAcronym{SDFEC}{\acroSCaps{sd-fec}}{soft-decision forward error correction}
\nAcronym{SDM}{\acroSCaps{sdm}}{space-division multiplexing}
\nAcronym{SE}{\acroSCaps{se}}{spectral efficiency}
\nAcronym{SER}{\acroSCaps{ser}}{symbol error rate}
\nAcronym{SI}{\acroSCaps{si}}{step index}
\nAcronym{SIFMF}{\acroSCaps{si-fmf}}{\usuk{step-index few-mode fiber}{step-index few-mode fibre}}
\nAcronym{SISMF}{\acroSCaps{si-smf}}{\usuk{step-index single-mode fiber}{step-index single-mode fibre}}
\nAcronym{SLM}{\acroSCaps{slm}}{spatial light modulator}
\nAcronym{SKR}{\acroSCaps{skr}}{secret key rate}
\nAcronym{SMD}{\acroSCaps{smd}}{spatial-mode dispersion}
\nAcronym{SSC}{\acroSCaps{ssc}}{spot-size converter}
\nAcronym{SMF}{\acroSCaps{smf}}{\usuk{single-mode fiber}{single-mode fibre}}
\nAcronym{SMU}{\acroSCaps{smu}}{source measure unit}
\nAcronym{SMUX}{\acroSCaps{smux}}{spatial multiplexer}
\nAcronym{SNR}{\acroSCaps{snr}}{signal-to-noise ratio}
\nAcronym{SNU}{\acroSCaps{snu}}{shot-noise unit}
\nAcronym{SOA}{\acroSCaps{soa}}{semiconductor optical amplifier}
\nAcronym[firstplural=\usuk{states of polarization (\acroSCaps{sop})}{states of polarisation (\acroSCaps{sop})}]{SOP}{\acroSCaps{sop}}{\usuk{state of polarization}{state of polarisation}}
\nAcronym{SPM}{\acroSCaps{spm}}{self-phase modulation}
\nAcronym{SPD}{\acroSCaps{spd}}{single-photon detector}
\nAcronym{SPS}{\acroSCaps{sps}}{samples per symbol}
\nAcronym{SRS}{\acroSCaps{srs}}{stimulated Raman scattering}
\nAcronym{SSB}{\acroSCaps{ssb}}{single-sideband}
\nAcronym{SSBI}{\acroSCaps{ssbi}}{signal-signal beat interference}
\nAcronym{SSFM}{\acroSCaps{ssfm}}{split-step Fourier method}
\nAcronym{SSMF}{\acroSCaps{ssmf}}{\usuk{standard single-mode fiber}{standard single-mode fibre}}
\nAcronym{STAXT}{\acroSCaps{staxt}}{short-term average cross-talk}
\nAcronym{STL}{\acroSCaps{stl}}{swept tunable laser}
\nAcronym{SVD}{\acroSCaps{svd}}{singular value decomposition}
\nAcronym{SW}{\acroSCaps{sw}}{sequence-wise}
\nAcronym{SWI}{\acroSCaps{swi}}{swept wavelength interferometry}
\nAcronym{SP}{\acroSCaps{sp}}{setpoint}

\nAcronym{TAT}{\acroSCaps{tat}}{transatlantic}
\nAcronym{TC}{\acroSCaps{tc}}{tunable coupler}
\nAcronym{TD}{\acroSCaps{td}}{time domain}
\nAcronym{TDE}{\acroSCaps{tde}}{\usuk{time domain equalizer}{time domain equaliser}}
\nAcronym{TDFA}{\acroSCaps{tdfa}}{thulium doped-fiber amplifier}
\nAcronym{TDM}{\acroSCaps{tdm}}{time-domain multiplexing}
\nAcronym{TDMSDM}{\acroSCaps{tdm-sdm}}{time-domain multiplexed space-division multiplexing}
\nAcronym{TE}{\acroSCaps{te}}{transverse electric}
\nAcronym{TRNG}{\acroSCaps{TRNG}}{true random number generator}
\nAcronym{TEC}{\acroSCaps{tec}}{thermally-expanded-core}
\nAcronym{TOPS}{\acroSCaps{tops}}{thermo-optic phase shifter}
\nAcronym{TLS}{\acroSCaps{tls}}{tunable laser source}
\nAcronym{TIA}{\acroSCaps{tia}}{trans-impedance amplifier}
\nAcronym{TM}{\acroSCaps{tm}}{transverse magnetic}
\nAcronym{TH4D}{\acroSCaps{th-4d}}{time domain hybrid four-dimensional}
\nAcronym{TH4D2A8PSK}{\acroSCaps{th-4d-2a8psk}}{time-domain hybrid four-dimensional two-amplitude eight-phase-shift keying}
\nAcronym{TP}{\acroSCaps{tp}}{twisted pair}
\nAcronym{TTL}{\acroSCaps{ttl}}{transistor-transistor logic}

\nAcronym{UWB}{\acroSCaps{uwb}}{ultra-wideband}
\nAcronym{ULI}{\acroSCaps{uli}}{ultrafast laser inscription}

\nAcronym{VCSEL}{\acroSCaps{vcsel}}{vertical-cavity surface emitting laser}
\nAcronym{VHDL}{\acroSCaps{vhdl}}{\acroSCaps{Vhsic} Hardware Description Language}
\nAcronym{VOA}{\acroSCaps{voa}}{variable optical attenuator}

\nAcronym{WDL}{\acroSCaps{wdl}}{wavelength-dependent loss}
\nAcronym{WDM}{\acroSCaps{wdm}}{wavelength-division multiplexing}
\nAcronym{WFS}{\acroSCaps{wfs}}{wavefront sensor}
\nAcronym{WGA}{\acroSCaps{wga}}{weakly guiding approximation}
\nAcronym{WGN}{\acroSCaps{wgn}}{white Gaussian noise}
\nAcronym[longplural=wavelength selective switches]{WSS}{\acroSCaps{wss}}{wavelength selective switch}
\nAcronym{WC}{\acroSCaps{wc}}{weakly coupled}
\nAcronym{WCMCF}{\acroSCaps{wc-mcf}}{\usuk{weakly-coupled multi-core fiber}{weakly-coupled multi-core fibre}}

\nAcronym{XGM}{\acroSCaps{xgm}}{cross-gain modulation}
\nAcronym{XPM}{\acroSCaps{xpm}}{cross-phase modulation}
\nAcronym{XOR}{\acroSCaps{xor}}{exclusive or}
\langcheck{%
    \nAcronym{XPOLM}{\acroSCaps{xp}ol\acroSCaps{m}}{cross-polarization modulation}
    }{%
    \nAcronym{XPOLM}{\acroSCaps{xp}ol\acroSCaps{m}}{cross-polarisation modulation}
}%
\nAcronym{XT}{\acroSCaps{xt}}{cross-talk}

\nAcronym{3DWG}{\acroSCaps{3dwg}}{3D-waveguide}

\titlespacing{\section}{0pt}{10pt}{5pt}
\usepackage{amsmath,amssymb}
\usepackage[style=ieee,sorting=none,maxnames=1,minnames=1]{biblatex}
\usepackage{float}
\usepackage{stfloats}
\begin{document}

\title{Characterising Interleaver Length in a Turbulent Deployed Terrestrial Free-Space Optical Link
\vspace{-8mm}
}

\author{
Kadir G\"{u}m\"{u}\c{s},
Vincent van Vliet,
Menno van den Hout,
Thomas Bradley,
Eduward Tangdiongga,
Chigo Okonkwo
\\
Electro-Optical Communication Group, Eindhoven University of Technology, The Netherlands, \textcolor{blue}{\url{k.gumus@tue.nl}}
}

\IEEEaftertitletext{\vspace{-5.2mm}}
\maketitle

\begin{abstract}
We investigate the correlation between interleaver length and turbulence for a deployed terrestrial free-space optical link spanning 4.6 km. Using 24 hours of measurements, we show an indicative distribution of interleaver length for capacities up to 350 Gb/s/pol and outage probabilities down to $10^{-4}$.
\end{abstract}

\IEEEpubidadjcol
\begin{figure*}[b!]
    \vspace{-5mm}
    \centering
    \resizebox{\linewidth}{!}{\input{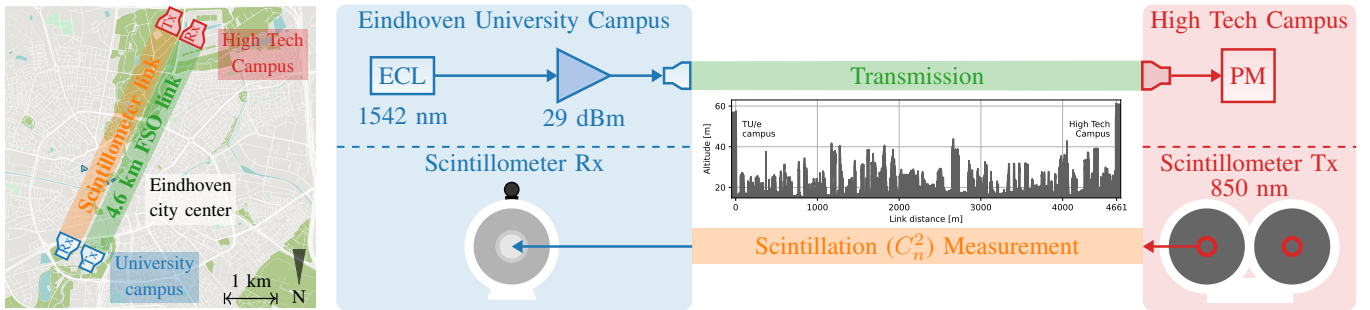}}
    \caption{Experimental setup for power and scintillation measurements for the Reid Photonloop.}
    \label{fig:setup}
\end{figure*}

\begin{figure*}[t!]
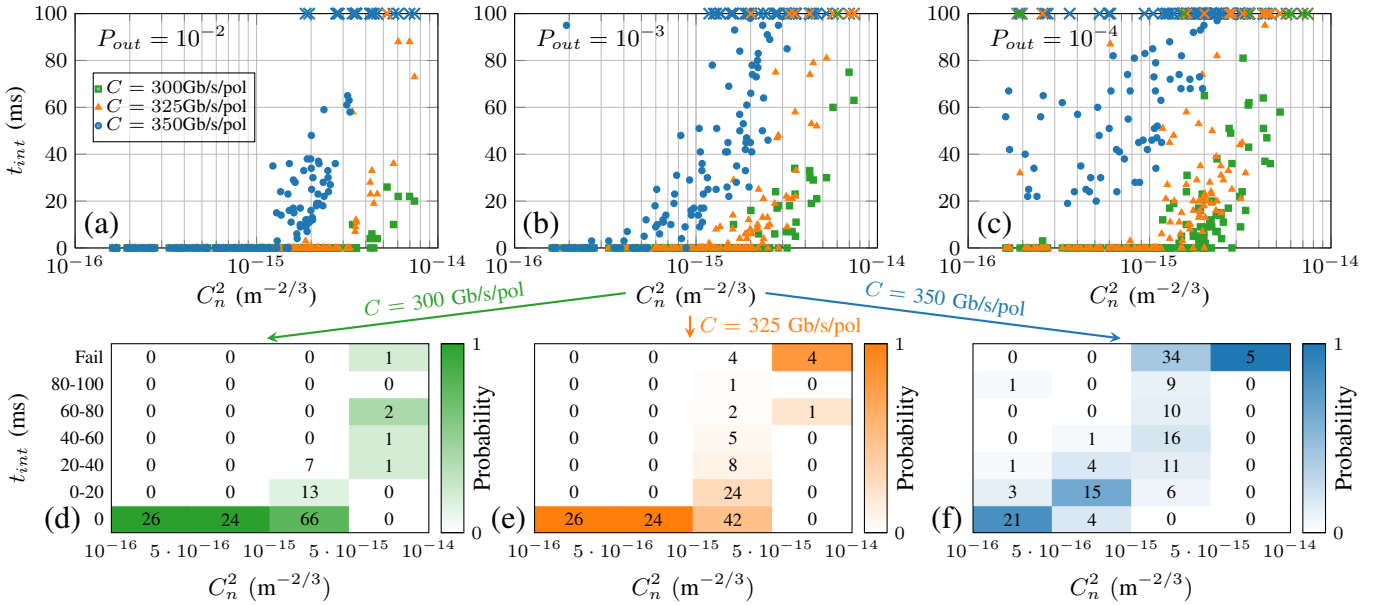

\begin{tikzpicture}
    \node at (0,0){\input{tikz/Interleaver_CN2_Pout2}};
    \node at (6,0){\input{tikz/Interleaver_CN2_Pout3}};
    \node at (12,0){\input{tikz/Interleaver_CN2_Pout4}};

    \node at (0.2,-4.13){\input{tikz/Interleaver_300}};
    \node at (6.3,-4.13){\input{tikz/Interleaver_325}};
    \node at (12.1,-4.13){\input{tikz/Interleaver_350}};

    \draw[thick,C2,-{stealth}] (6,-2.1) -- node[font = \footnotesize,xshift = 12mm]{$C = $ 325 Gb/s/pol} (6,-2.4);
    \draw[thick,C3,-{stealth}] (5.15,-1.8) -- node[font = \footnotesize, yshift = 2mm, sloped]{$C = $ 300 Gb/s/pol}(0.4,-2.4);
    \draw[thick,C1,-{stealth}] (7,-1.8) -- node[font = \footnotesize, yshift = 2mm, sloped]{$C = $ 350 Gb/s/pol}(11.7,-2.4);

    \node[font = \large] at (-2.3,-4.8){(d)};
    \node[font = \large] at (3.6,-4.8){(e)};
    \node[font = \large] at (9.4,-4.8){(f)};
\end{tikzpicture}
\caption{(a,b,c): $t_{int}$ vs. $C_n^2$ for different target $P_{out}$ and target $C$. The crosses indicate points at which either an interleaver longer than 100ms is required or infinite interleaver capacity is too low. (d,e,f): Distribution of $t_{int}$ for $P_{out} = 10^{-3}$ for 3 different target capacities. The probabilities are normalised per column; the numbers in each cell correspond to the total number of occurrences.
\vspace{-2mm}
}
\label{fig:results}
\end{figure*}

\vspace{-2mm}
\section{Introduction}
\vspace{-2mm}
Terrestrial \FSO communication allows for wireless communication with high data rates by using coherent optical fibre transmission technologies. Terrestrial links can be used for applications such as emergency networks, quantum key distribution, and cellular backhaul. In recent years, high-capacity terrestrial \FSO links have been experimentally demonstrated \cite{vanVliet_JLT,bai2025112}, showcasing their viability. However, \FSO links suffer from reliability issues because of atmospheric turbulence; the received optical power, and therefore the \SNR of the system, fluctuates significantly \cite{khalighi2014survey}. For terrestrial links, these fluctuations are often more severe than for satellite links, as the beam must travel farther through the dense part of the atmosphere. To improve system stability, data interleaving for \FSO systems has been proposed \cite{shi2004interleaving}. Interleaving spreads burst errors caused by fading over the interleaver time. This averages the capacity over the interleaving period, increasing system stability at the cost of higher latency. While interleaving strategies have been extensively investigated for satellite FSO systems \cite{Ollie2026}, studies on terrestrial FSO links remain limited \cite{Greco2009,fujita2019experimental}. Existing terrestrial studies do not establish a quantitative relationship between atmospheric turbulence conditions and the interleaver lengths required to achieve specific reliability targets. As a result, there is currently no clear framework for adaptively configuring interleavers in practical terrestrial FSO deployments.

In this work, we address this gap by experimentally investigating the correlation between interleaver length $ t_{int}$ and the refractive index structure parameter $C_n^2$ using 24 hours of measurement data from the Reid Photonloop testbed \cite{vanVliet_JLT}, a terrestrial \FSO link spanning 4.6 km across the city of Eindhoven. We show an indicative distribution of $t_{int}$ for capacities up to 350 Gb/s/pol and outage probabilities down to $10^{-4}$. These results provide practical design guidelines for adaptive interleaving in future terrestrial \FSO systems.

\vspace{-2mm}
\section{Experimental Setup and Results}
\vspace{-2mm}
Fig. \ref{fig:setup} shows the Reid Photonloop testbed, a terrestrial \FSO setup spanning 4.6~km across the city of Eindhoven. For the investigated \FSO channel, the transmitter is located on the Eindhoven University of Technology campus, and the receiver is located on the High Tech Campus. At the transmitter side, we use an \ECL at 1542~nm, which is amplified to 29~dBm by an \EDFA. Using fibre-coupled optical terminals designed by Aircision, the light is transmitted across the city.  These terminals are exemplative of common terminals used in terrestrial \FSO systems \cite{vanVliet_JLT}. Using a fibre-coupled, high-speed power meter, we measure the received power $P_{rec}$ every 10 \SI{}{\micro\text{s}} at the receiver side. In parallel with the power measurements, we measure the refractive index structure parameter $C_n^2$, which indicates turbulence strength. For this, we use the Scintec BLS900 NEO Dual-Disk Design scintillometer, which characterises $C_n^2$ with an averaging time of 10 minutes, with the transmitter at the High Tech Campus and the receiver at the Eindhoven University of Technology campus. The measurements were performed on October 18$^{\text{th}}$ 2025 throughout the day. 

We assume a receiver with an \EDFA and a noise figure of 4.4~dB, operating over a bandwidth $W$ of 50~GHz. Using eq. (54) from \cite{essiambre2010capacity} we calculate the ASE noise generated by the \EDFA. The total \SNR of the system is defined as $\text{SNR} = \frac{1}{1/\text{SNR}_{B2B} + 1/\text{SNR}_{EDFA}}$, where $\text{SNR}_{B2B}$ is the \SNR of our back-to-back system measured at 22~dB, and $\text{SNR}_{EDFA}$ is the \SNR of the shot-noise-limited signal amplified by an \EDFA. The capacity for a single polarisation is then given by $C = W\log_2(1+\text{SNR})$. For our system, when using an infinitely long interleaver, $C$ peaks at around 360 Gb/s/pol.

The power meter measurement data were split into one-minute segments to calculate the capacity after interleaving. We ignore any minutes during which we are unable to take measurements due to a power meter failure, or when $P_{rec}$ measurements were unusually low ($P_{rec} < -70$ dBm), most likely caused by external interference or failure of the alignment control loop. We also filtered out one 10-minute segment because no scintillation measurement was recorded during that time. Over the 1440 measurement minutes, 1328 were used for analysis. To calculate the interleaver capacity for an ideal interleaver with $t_{int} \leq 100$~ms, we apply sliding-window averaging to the non-interleaved capacity. The outage probability $P_{out}$ is the probability that the system's capacity is lower than the target capacity. We calculate this outage probability for each 10-minute scintillation measurement, yielding 141 data points, since three 10-minute segments were completely filtered out. 

In Fig. \ref{fig:results} (a,b,c), we show $t_{int}$ vs $C_n^2$ for target $P_{out} = \{10^{-2},10^{-3},10^{-4}\}$ and target $C = \{300,325,350\}$ Gb/s/pol. A clear correlation between $t_{int}$ and $C_n^2$ is evident, with $t_{int}$ increasing as $C_n^2$, hence the turbulence, increases. Similarly, when $P_{out}$ decreases, longer interleavers are required. For $P_{out} = 10^{-3}$, we see an outlier at $C_n^2 = 2\cdot10^{-16}$ when $C = 350$ Gb/s/pol, caused by an unusually long fade. The distribution of $t_{int}$ is shown in Fig. \ref{fig:results} (d,e,f) for $P_{out} = 10^{-3}$ for the three different target capacities. For $C_n^2 < 10^{-15}$, no interleaving is required for $C \leq 325$ Gb/s/pol, while for $C = 350$ Gb/s/pol the spread of the distribution is small, with the exception of the aforementioned outlier. Most of our scintillation measurements were for $10^{-15} \leq C_n^2 \leq 5\cdot10^{-15}$, where for $C \geq 325$ Gb/s/pol, the spread of required $t_{int}$ is very wide, with the target outage probability not always being reached. This shows that for this range of $C_n^2$, more than just scintillation data is required for choosing $t_{int}$. For $C_n^2 > 5\cdot 10^{-15}$, the target capacity needs to be reduced significantly to allow for low $P_{out}$, as the interleaver is not able to handle all of the fading. Although these results are for our \FSO system over a single day, the analysis and insights gained from them are applicable to other terrestrial \FSO systems.

\vspace{-2mm}
\section{Conclusions}
\vspace{-2mm}

We analysed the distribution of required interleaver lengths for a deployed 4.6 km terrestrial FSO link, considering capacities up to 350 Gbit/s/pol and outage probabilities down to $10^{-4}$. Using 24 hours of experimental data, we demonstrated a strong correlation between the refractive index structure parameter $C_n^2$ and interleaver length. Higher turbulence levels were found to increase both the required interleaver length and its variability. These results provide a basis for dimensioning interleavers in terrestrial FSO systems and pave the way toward adaptive interleaving schemes that dynamically balance reliability and latency under varying atmospheric conditions.

\vspace{2mm}
\scriptsize \noindent
Supported by the Dutch Research Council (NWO) TTW-Perspectief Optical Wireless Superhighways: Free photons (at home and in space): FREE P19-13, the Dutch Ministry of Economic Affairs and Climate Policy (EZK) via the PhotonDelta National Growth Fund Programme on Photonics, and European Innovation Council Transition project CombTools (G.A. 101136978). We thank Aircision B.V., particularly Nourdin Kaai, Luis Pellicer Collado, and Roland Blok, for their support of the Reid Photonloop \FSO testbed, and Keysight Technologies for providing the high-speed power meter.
\vspace{-2mm}
\printbibliography

@article{essiambre2010capacity,
  title={Capacity limits of optical fiber networks},
  author={Essiambre, Ren{\'e}-Jean and Kramer, Gerhard and Winzer, Peter J and Foschini, Gerard J and Goebel, Bernhard},
  journal={Journal of Lightwave technology},
  volume={28},
  number={4},
  pages={662--701},
  year={2010},
  publisher={IEEE},
}

@ARTICLE{vanVliet_JLT,
  author={van Vliet, Vincent and van den Hout, Menno and G\"{u}m\"{u}\c{s}, Kadir and Tangdiongga, Eduward and Okonkwo, Chigo},
  journal={Journal of Lightwave Technology}, 
  title={{Multi-Terabit Coherent Free-Space Optical Communication Over a 4.6 km Urban Channel}}, 
  year={2026},
  volume={},
  number={},
  pages={1-9},
}

@article{bai2025112,
  title={112 {G}bit/s single-wavelength {FSO} communication with 104.8 km horizontal atmospheric link over {Q}inghai {L}ake},
  author={Bai, Zhaofeng and Bian, Yiming and Wang, Xuan and Chang, Yidi and Wang, Xiong and Quan, Xinxin and Li, Yuang and Li, Ke and Liu, Huan and Gao, Duorui and others},
  journal={Optics Express},
  volume={33},
  number={9},
  pages={19966--19979},
  year={2025},
  publisher={Optica Publishing Group},
}

@article{khalighi2014survey,
  title={Survey on free space optical communication: {A} communication theory perspective},
  author={Khalighi, Mohammad Ali and Uysal, Murat},
  journal={IEEE communications surveys \& tutorials},
  volume={16},
  number={4},
  pages={2231--2258},
  year={2014},
  publisher={IEEE},
}

@article{shi2004interleaving,
  title={Interleaving for combating bursts of errors},
  author={Shi, Yun Q and Zhang, Xi Min and Ni, Zhi-Cheng and Ansari, Nirwan},
  journal={IEEE circuits and systems magazine},
  volume={4},
  number={1},
  pages={29--42},
  year={2004},
  publisher={IEEE},
}

@inproceedings{Ollie2026,
author = {Farley, Ollie and Lognoné, Perrine and Boddeda, Rajiv and Osborn, James},
booktitle = {Optical Fiber Communication Conference (OFC) 2026},
publisher = {Optica Publishing Group},
title = {Leveraging Network Diversity for Capacity Maximization in European Optical {GEO} Feeder Systems under Realistic Optical Turbulence},
year = {2026},
}

@inproceedings{Greco2009,
author = {Joseph A. Greco},
title = {{Design of the high-speed framing, FEC, and interleaving hardware used in a 5.4km free-space optical communication experiment}},
volume = {7464},
booktitle = {Free-Space Laser Communications IX},
publisher = {SPIE},
pages = {746409},
year = {2009},
}

@inproceedings{fujita2019experimental,
  title={Experimental evaluation of polar code transmission in terrestrial free-space optics},
  author={Fujita, Shingo and Ito, Keita and Okamoto, Eiji and Takenaka, Hideki and Kunimori, Hiroo and Endo, Hiroyuki and Fujiwara, Mikio and Kitamura, Mitsuo and Sasaki, Masahide and Toyoshima, Morio and others},
  booktitle={2019 IEEE International Conference on Space Optical Systems and Applications (ICSOS)},
  pages={1--6},
  year={2019},
  organization={IEEE},
}
\end{document}